\documentclass{article}
\usepackage{fullpage}
\usepackage{amsmath}
\usepackage{amsfonts}
\usepackage{graphicx, color, tabularx}

\renewcommand{\baselinestretch}{1.25}

\newtheorem{lem}{Lemma}

\newtheorem{thm}{Theorem}

\newcolumntype{S}{>{\centering\arraybackslash} m{.4\linewidth} }

\begin{document}
\bibliographystyle{IEEEtran}
\title{Normal Factor Graphs and Holographic Transformations}
\author{Ali Al-Bashabsheh and Yongyi Mao\\ University of Ottawa\\ School of Information Technology and Engineering}
\date{}
\maketitle

\begin{abstract}
This
paper stands at the intersection of two distinct lines of research.
One line is ``holographic algorithms,'' a powerful approach introduced by Valiant for solving
various counting problems in computer science; the other is ``normal factor graphs,'' an elegant framework proposed by Forney for representing codes defined on graphs. We introduce the notion of holographic transformations for normal factor graphs, and establish a very general theorem, called the generalized Holant theorem, which relates a normal factor graph to its holographic transformation. We show that the generalized Holant theorem on the one hand underlies the principle of holographic algorithms, and on the other hand reduces to a general duality theorem for normal factor graphs, a special case of which was first proved by Forney. In the course of our development, we formalize a new semantics for normal factor graphs, which highlights various linear algebraic properties that potentially enable the use of normal factor graphs as a linear algebraic tool.
\end{abstract}

\section{Introduction}

The formulation of codes on graphs and the invention of iterative decoding algorithms for graphical codes have undoubtedly revolutionized  coding theory. In this research area, the introduction of normal graphs \cite{Forney2001:Normal} and their duality properties are arguably one of the most elegant and profound results.

In his celebrated paper \cite{Forney2001:Normal}, Forney introduced the notion of {\em normal realizations}, represented by {\em normal graphs}, as generalized state realizations of codes. In a normal graph, each vertex represents a local (group) code constraint, and each edge represents a variable that is involved in either one or two of the constraints. A variable involved in two constraints, called a {\em state} variable, is represented by a regular edge, namely, an edge connecting two vertices; the two vertices correspond
to the two constraints involving the variable. A variable involved in only one constraint, called a {\em symbol} variable, is represented by a ``dangling edge,''\footnote{In Forney \cite{Forney2001:Normal}, such edges are called ``half-edges''.} namely, an edge incident on only one vertex;  the vertex corresponds to the single constraint involving the variable.
The {\em global behavior} represented by the normal graph is the set of all symbol-state configurations satisfying all local constraints,
and the realized code is the set of all symbol configurations that participate in at least one symbol-state configuration in the global behavior. Forney showed in \cite{Forney2001:Normal} that  codes realized by Tanner graphs \cite{Tanner81} or Wiberg-type graphs \cite{TWL:Wiberg, TWL:WibergEtal}, in which 
a variable may in general be involved in an unrestricted number of constraints,
can be converted to normal graphs by properly replicating some variables.

 Normal realizations of codes have a fundamental duality property. By introducing a simple local ``dualization'' procedure that converts each local code in a normal realization to its dual code and inserts additional ``sign inverters,'' Forney proved a {\em normal graph duality theorem} 
\cite{Forney2001:Normal}, which shows that the dualized normal graph realizes the dual code. 
 
 The notion of a normal graph may be extended to the notion of a {\em normal factor graph}, or {\em Forney-style factor graph} \cite{Loeliger:Intro, Loeliger:FG}, which uses an identical graphical representation, but in which the graph vertices no longer represent constraints. Instead, in a normal factor graph, each vertex represents a local {\em function} involving precisely the variables represented by the edges incident to the vertex. Treated as a factor graph \cite{frank:factor} with particular variable-degree restrictions, and interpreted using the
standard semantics of factor graphs, a normal factor graph represents a multivariate function that factors as the product of all of the local functions that are represented by the graph vertices. When each local function in the normal factor graph is the indicator function of a local code constraint, the represented function is the indicator function of the global behavior. This makes normal factor graph representations of codes equivalent to normal graphs, and allows the translation of the normal graph duality theorem to an equivalent theorem for normal factor graphs that represent codes.


 

Among the first to recognize the profound value of normal graphs, Koetter \cite{Koetter2002:1} applied normal graphs (or normal factor graphs for codes) and the duality theorem in a study of trellis formations \cite{Koetter:Formation1}, which gives a necessary and sufficient condition of mergeability and a polynomial-time algorithm for deciding whether a trellis formation contains mergeable vertices. In addition, Koetter {\em et al}.  \cite{Koetter2004:1} extended the applications of normal graphs from channel codes to network codes, and used the fundamental duality theorem  to establish a reversibility theorem in network coding.

In a seemingly distant research area of complexity theory, Valiant proved the tractability of families of combinatorial problems for which no polynomial-time solvers were known previously in a ground-breaking  work \cite{Valiant2004:Holographic}. In this paper, Valiant develops a very powerful family of algorithms, 
which he calls \emph{holographic algorithms}, to solve such problems. Holographic algorithms are based on the concept of ``holographic reduction,'' and are governed by a fundamental theorem that Valiant calles the {\em Holant theorem}.

Although in his original work \cite{Valiant2004:Holographic} Valiant dealt only with transforming a product of functions to a specific form, the Holant theorem 
establishes a principle for transforming an arbitrary product of functions to another product of functions
such that the sum over the configuration space is unchanged.
Consequently, when computing the sum of a product of functions, the Holant theorem provides a family of transformations that convert the product to a different one, for which the sum may be efficiently computable. 

Since many problems in coding and information theory require the computation of sums of products (such as in decoding error correction codes and in computing certain capacities), holographic algorithms become a potentially powerful tool for the information theory community. Indeed, in \cite{Schwartz:2008}, Schwartz and Bruck showed that certain constrained-coding capacity problems may be solved in polynomial time using holographic algorithms.

This paper stands at the intersection of the above-mentioned two lines of research, bridging the two areas by unifying Valiant's holographic reduction and Forney's normal-graph dualization with the notion of ``holographic transformations,''  a term that we coin in this paper. The focus of this paper will be on general normal factor graphs, in which the vertices may represent arbitrary functions rather than just indicator functions. 

We first introduce a new semantics for normal factor graphs, which we call the ``exterior-function semantics." In the
exterior-function semantics, instead of letting the graph represent a product of the local functions, we let it represent a sum of products of local functions, which we call the ``exterior function."  In this setting, a normal factor graph may be viewed as an expression, or realization, of its exterior function in terms of a ``sum of products." In fact, the notion of ``sum of products'' is the key connection between Forney's normal graph duality theorem and Valiant's Holant theorem: in the case of Valiant \cite{Valiant2004:Holographic}, this ``sum of products'' is the number of configurations that needs to be computed, and in the case of Forney \cite{Forney2001:Normal} (when using normal factor graphs rather than normal graphs to represent codes), this ``sum of products'' is a code indicator function (possibly up to a scale factor). In this new framework, a holographic transformation is defined as a transformation of a normal factor graph that changes all local functions subject to certain conditions, and converts the normal factor graph to a structurally identical one.

The main result of this paper is what we call the {\em generalized Holant theorem},
relating a normal factor graph and its holographic transformations in terms of their realized functions.
On the one hand, we show that the Holant theorem in \cite{Valiant2004:Holographic} is a special case of the generalized Holant theorem. 
On the other hand, we prove a general duality theorem for normal factor graphs as a corollary of the generalized Holant theorem. This duality theorem reduces to Forney's original normal graph duality theorem for normal factor graphs that represent codes.

Another result of our development is a new  understanding of normal factor graphs from a linear algebraic perspective. More specifically, the exterior-function semantics associates a normal factor graph with a ``sum-of-products'' form, and a sum-of-products form may be regarded as a linear algebraic expression such as a vector dot product, matrix product or tensor product. In contrast to conventional 
notations for algebraic expressions, in which re-ordering terms results in illegitimate or different mathematical expressions, sum-of-products forms and their normal factor graph representations eliminate the need to properly order terms (\emph{i.e.}, the local functions).  This allows a transparent development of our results. To us, normal factor graphs with the
 exterior-function semantics appear  to be a very natural and intuitive language for linear algebra, and may potentially be useful in a wide variety of applications.

The holographic transformations introduced in this paper equip normal factor graphs with a rich family of linear transformations, potentially enabling normal factor graphs to serve as a more general analytic framework and computational tool. The power of these transformations, in addition to providing a fundamental duality theorem in coding theory, has also  been hinted at by the great power of holographic algorithms (see, \emph{e.g.}, \cite{Valiant2004:Holographic, Schwartz:2008, Cai2007:Art, Cai2008:Fibonacci, Valiant:Observations2, Cai:MatchgatesPlanar}).

This paper was initially motivated by our recognition of the connection between Forney's normal graph duality theorem and Valiant's Holant theorem. Interactions with the editors of this paper (G. D. Forney, Jr.\! and P. Vontobel) and communications with 
Forney on his concurrent development of
 ``partition functions of normal factor graphs'' \cite{Forney:MacWilliams2} have also provided important inspirations for our development. Indeed, much of our development shares similarities with Forney's approach in \cite{Forney:MacWilliams2}, which also proves the general normal factor graph duality theorem.


During the review process, we learned from the reviewers and the editors  of this paper that the techniques used to establish our results are similar to those in some previous literature, including, for example, the notion of ``loop calculus'' in \cite{Chertkov:Loop1, Chertkov:Loop2, Chertkov:Loop3, Chertkov:easy}, and
the ``opening/closing the box'' approach in \cite{Pascal:Electrical, Pascal:Kalman, Loeliger:Intro}. We shall give proper credit to these authors as we proceed in this paper.

The remainder of the paper is structured as follows. Section \ref{sec:NFG} formalizes the exterior-function semantics for normal factor graphs, discusses its linear algebraic interpretations, introduces the notion of holographic transformations, and establishes the generalized Holant theorem. Section \ref{sec:application} discusses applications of holographic
transformations, including deriving the general normal factor graph duality
theorem as a corollary to generalized Holant theorem, and explaining the principle of holographic reduction.
 The paper is briefly concluded in Section \ref{sec:conclude}.

\section{Normal Factor Graphs and Holographic Transformations}
\label{sec:NFG}

The term ``normal factor graph'' has been used in the literature with various meanings. 
In particular, normal factor graphs as defined in  \cite{Forney2001:Normal} can be easily confused with normal graphs, normal factor graphs for codes, or graphs  in which variables  are represented by variable vertices.
Making a joint effort with Forney \cite{Forney:MacWilliams2}, we advocate in this paper a more rigorous use of the term ``normal factor graph,'' and introduce a new semantics, the ``exterior-function'' semantics,\footnote{The same semantics is also presented in the concurrent development of Forney \cite{Forney:MacWilliams2}, in which what we call
``exterior functions'' are called ``partition functions.''} that defines what a normal factor graph means. To us, this new semantics is quite appealing, since it allows clean development of various graph properties, and has an elegant linear algebraic perspective.

\subsection{Normal Factor Graphs: The Exterior-Function Semantics}

Formally, a {\em normal factor graph} (NFG) is a graph $(V, E)$, with vertex set $V$ and edge set $E$, where the edge set $E$ consists of two kinds of edges, a set $E^{\bf \rm int}$ of ordinary edges, each connecting two vertices, and a set $E^{\bf \rm ext}$ of
``dangling edges,'' each having one end attached to a vertex and the other end free.\footnote{More formally, such dangling edges are
hyperedges of degree 1, and thus strictly speaking a NFG is a hypergraph rather than a graph.}
Each edge $e\in E$ represents a variable $x_e$ taking values from some finite alphabet\footnote{It is possible to generalize the results of this paper to the cases where the alphabets are infinite or uncountable, although this involves some subtle technicalities.} ${\cal X}_e$; sometimes  we may alternatively say that the edge $e$ represents the 
alphabet ${\cal X}_e$ when we do not wish to specify the variable name.
Each vertex $v$ represents a complex-valued function\footnote{All functions
in this paper are complex-valued. However, all results can be generalized to $\mathbb{F}$-valued functions, where ${\mathbb F}$ is an arbitrary field.} $f_v$ on the cartesian product ${\cal X}_{E(v)}:=\prod\limits_{e\in E(v)} {\cal X}_e$, where $E(v)$ is the set of all edges incident to $v$. If we denote the set $\{f_v:v\in V\}$ of functions by $f_V$, then the NFG
is specified by the tuple $(V, E^{\rm int}, E^{\rm ext}, f_V)$.

When treated as a factor graph under the conventional semantics \cite{frank:factor}, the NFG ${\cal G}=(V, E^{\rm int}, E^{\rm ext}, f_V)$ 
represents the product $\prod\limits_{v\in V}f_v(x_{E(v)})$ of 
all functions in $f_V$.  This product, expressing a function on ${\cal X}_E:=\prod\limits_{e\in E}{\cal X}_e$, will be called the {\em interior function}\footnote{In the conventional factor graph literature, an interior function is referred to as a ``global function."} realized by the NFG. Here we have used the standard ``variable set'' notation $x_{E(v)}$ to denote the set of variables $\{x_e:e\in E(v)\}$.  

Now  we introduce a new semantics for NFGs: instead of letting an NFG 
represent its interior function, we let it represent the interior function summed over all
variables represented by regular edges.  We call this function
the {\em exterior function} realized by the NFG. More precisely, the exterior function realized by the NFG ${\cal G}=(V, E^{\rm int}, E^{\rm ext}, f_V)$ is the function
\begin{equation}
\label{eq:exteriorFunc}
Z_{\cal G}(x_{E^{\rm ext}}):= \sum\limits_{x_{E^{\rm int}}}\prod\limits_{v\in V}f_v(x_{E(v)}).
\end{equation}
Thus the exterior function involves only the external variables, represented by the dangling edges. Letting the NFG ${\cal G}$ express the function $Z_{\cal G}$ as
defined in (\ref{eq:exteriorFunc}) is what we call the {\em exterior-function semantics}, which we will use throughout this paper.

In this setting, one may view an NFG as an {\em expression}, or {\em realization}, of a function (the exterior function) that is given in ``sum-of-products'' form, as in (\ref{eq:exteriorFunc}), where each variable is involved in either one function or two functions, and
the summation is over all variables that are involved in two functions. Since a variable involved in two functions (represented by a regular edge) is ``invisible'' in the exterior function, we call it an {\em internal variable}.
In contrast, a variable involved only in one function (represented by a dangling edge) remains visible in the exterior function, and is called an {\em external variable}.

\begin{figure}
\centerline{
\scalebox{0.6}{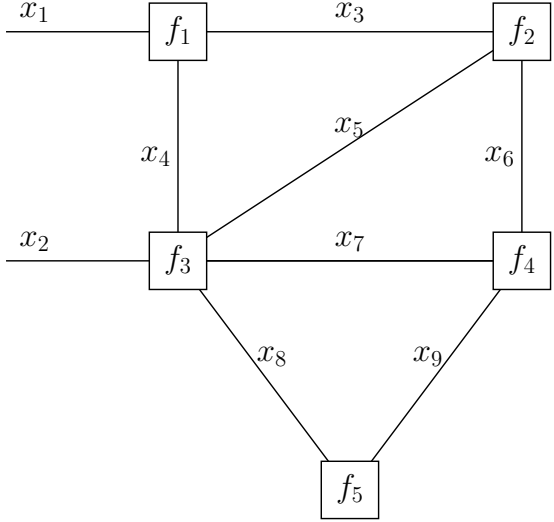}
}
\caption{A normal factor graph (NFG) ${\cal G}$.}
\label{fig:ex_NFG}
\end{figure}

An example of an NFG is given in Figure \ref{fig:ex_NFG}, which realizes the exterior function
\[
Z_{\cal G}(x_1, x_2)=\sum\limits_{x_3, \ldots, x_9} f_1(x_1, x_3, x_4) f_2(x_3, x_5, x_6) f_3(x_2, x_4, x_5, x_7, x_8) f_4(x_6, x_7, x_9)
f_5(x_8, x_9).
\]

At first glance,  NFGs and this semantics may appear to impose a restriction on which sum-of-products forms are representable. We note, however, that  {\em any} sum-of-products form can be straightforwardly converted to one that directly corresponds to an NFG. This requires only that we properly replicate variables, using
a ``normalization" procedure similar to that of Forney in \cite{Forney2001:Normal} for converting a factor graph to a normal graph. (The Appendix gives a detailed account of this procedure.) For this reason, using NFGs to represent sum-of-products forms entails no loss of expressive power.

For notational convenience, we may denote the sum-of-products form in (\ref{eq:exteriorFunc}) by\footnote{In the case of two arguments, say functions $f$ and $g$, the sum-of-products form $\langle f, g\rangle$ should not be confused with the Hermitian inner product of $f$ and $g$, whose definition requires complex conjugation of one of the two arguments.}
\[\big\langle f_1(x_{E(1)}), f_2(x_{E(2)}), \ldots, f_{|V|}(x_{E(|V|)})\big\rangle,\]
if $V$ is identified with the set $\{1, 2, \ldots, |V|\}$.
Due to the commutativity of both multiplication and summation and the distributive law relating the two operations,
it is easy to see that any ordering of the arguments of
$\langle \cdot , \cdot , \ldots , \cdot \rangle$  expresses the same function. Consequently, we may write the sum-of-products form more  compactly
as $\langle f_{v}(x_{E(v)}): v\in V\rangle$, or even as $\langle f_{v}:v\in V\rangle$, if no ambiguity results.

\subsection{Exterior-Function-Preserving Procedures}
\label{subsec:procedures}

The exterior-function semantics of NFGs allows us to identify immediately several elementary graph manipulation procedures that preserve the
exterior function.

\begin{figure}
\centerline{
\begin{tabular}{c@{\hspace{2cm}}c}
\scalebox{0.7}{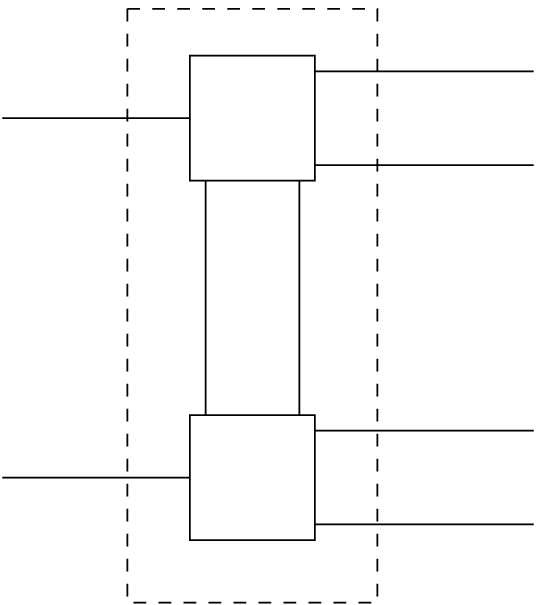}
&
\scalebox{0.7}{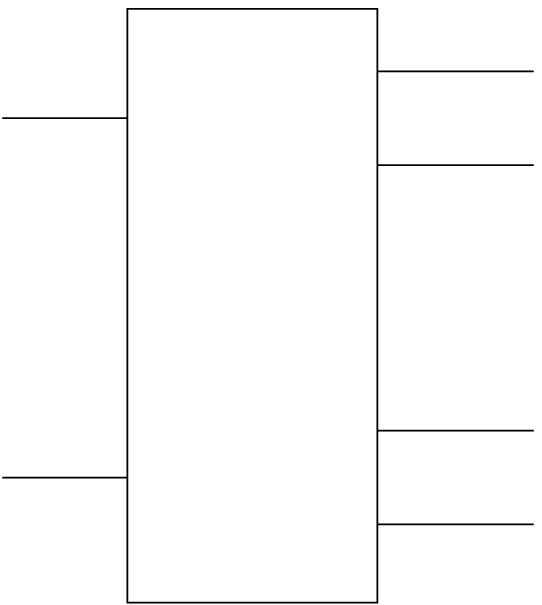}\\
& \\
\end{tabular}
}
\caption{Vertex Grouping/Splitting Procedure. Left to right: Vertex Grouping; right to left: Vertex Splitting.}
\label{fig:vertexGroupSplit}
\end{figure}

\vspace{0.3cm}
\noindent{\bf Vertex Grouping/Splitting Procedure} In a Vertex Grouping Procedure, two vertices representing functions $f$ and $g$ are grouped together, and the group is replaced by a vertex representing the function $\langle f, g\rangle$. In a Vertex Splitting Procedure, a vertex representing a function that can be expressed by the sum-of-products form $\langle f, g\rangle$ is replaced by an NFG representing $\langle f, g\rangle$. Figure \ref{fig:vertexGroupSplit} shows this pair of procedures.

\begin{lem}
\label{lem:vertexGroupSplit}
Applying a Vertex Grouping or Vertex Splitting Procedure to an NFG preserves the realized exterior function.
\end{lem}
\noindent{\em Proof:}
This lemma holds because these procedures simply correspond to conversions between sum-of-products forms
$\langle f, g, h_1, \ldots, h_m\rangle$ and $\big\langle \langle f, g\rangle, h_1, h_2, \ldots, h_m\big\rangle$, which evidently express the
same function. \hfill $\Box$

We note that this pair of procedures were first introduced by Loeliger in \cite{Loeliger:Intro, Loeliger:FG}, who refers to Vertex Grouping as
``closing the box,'' and to Vertex Splitting as ``opening the box." However, in \cite{Loeliger:Intro, Loeliger:FG}
these procedures are used to interpret an NFG (in the original semantics) as a flexible hierarchical model, and to explain message-passing algorithms, rather than in the context of the exterior-function semantics.

Note that the Vertex Grouping  Procedure may be applied recursively to an arbitrary number of  vertices, say $f_1, f_2, \ldots, f_m$, so that these functions are replaced by a single function realized by $\langle f_1, f_2, \ldots, f_m\rangle$. The resulting NFG still realizes the same exterior function. Similarly, the reverse Vertex Splitting Procedure also preserves the exterior function. For these reasons, when we draw a dashed box (as in Figure \ref{fig:vertexGroupSplit}, left) to group some vertices, we may freely interpret the NFG 
as the equivalent NFG in which the box is replaced by a single vertex representing the function realized by the box.

\begin{figure}[ht!]
\centerline{
\begin{tabular}{c@{\hspace{2cm}}c}
\scalebox{0.7}{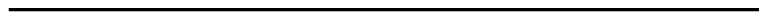}
&
\scalebox{0.7}{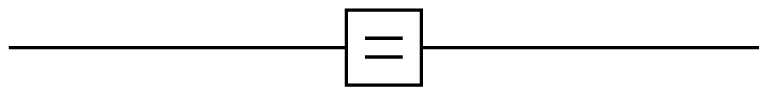}\\
&\\
\end{tabular}
}
\caption{Equality Insertion/Deletion Procedure.
Left to right: Equality Insertion; right to left: Equality Deletion. The edges may be regular or dangling; the vertex labelled with ``$=$'' represents the function $\delta_{=}$.}
\label{fig:eqInsertDelete}
\end{figure}

\vspace{0.3cm}
\noindent{\bf Equality Insertion/Deletion Procedure} For any finite alphabet ${\cal X}$, let $\delta_{=}$ denote the $\{0, 1\}$-valued function on ${\cal X}\times {\cal X}$ which evaluates to $1$ if and only if the two arguments of the function are equal. That is, $\delta_{=}$ is an ``equality indicator function." In an Equality Insertion Procedure, a $\delta_{=}$ function is inserted into an edge; in an Equality Deletion Procedure, a $\delta_{=}$ is deleted and the two edges originally connected to the function are joined. Figure \ref{fig:eqInsertDelete} shows this pair of procedures.

\begin{lem}
\label{lem:eqInsertDelete}
Applying an Equality Insertion or Deletion Procedure to an NFG preserves the realized exterior function.
\end{lem}
\noindent {\em Proof:}
This result is a corollary of Lemma \ref{lem:vertexGroupSplit}, and simply follows from the fact that if the $\delta_{=}$ function is inserted into an edge incident to a function $f$, then we may group $f$ with $\delta_{=}$ and replace the sum-of-products form $\langle f, \delta_{=}\rangle$ with the function it expresses, namely $f$. \hfill $\Box$

\vspace{0.3cm}
\noindent{\bf Dual Vertex Insertion/Deletion Procedure} Let ${\cal X}$ and ${\cal Y}$ be two finite alphabets and $\Phi:{\cal X}\times {\cal Y} \rightarrow {\mathbb C}$ and $\widehat{\Phi}:{\cal X}\times {\cal Y} \rightarrow {\mathbb C}$ be two functions. Then we say that $\Phi$ and $\widehat{\Phi}$ are  {\em dual} with respect to the alphabet ${\cal Y}$, and call ${\cal Y}$
the {\em coupling} alphabet,
if $\big\langle \Phi(x, y), \widehat{\Phi}(x', y)\big\rangle =\delta_{=}(x, x')$ for every  $x, x'\in {\cal X}$.  In the case when ${\cal X}$ and ${\cal Y}$ have the same cardinality, we also call the vertices representing $\Phi$ and $\widehat{\Phi}$ {\em transformers}, a term which will be justified in Section \ref{subsec:algebra}. In a Dual Vertex Insertion Procedure, we insert into an edge representing alphabet ${\cal X}$ a dual pair of functions $\Phi:{\cal X}\times {\cal Y}\rightarrow {\mathbb C}$ and $\widehat{\Phi}:{\cal X}\times {\cal Y}\rightarrow {\mathbb C}$, with  ${\cal Y}$ being the coupling alphabet, and let the edge connecting the two functions represent ${\cal Y}$. A Dual Vertex Deletion Procedure is the reverse of a Dual Vertex Insertion Procedure, in which we delete a pair of dual functions and the edge connecting them, and then join the ends of the two cut edges.

\begin{figure}
\centerline{
\begin{tabular}{c@{\hspace{2cm}}c}
\scalebox{0.7}{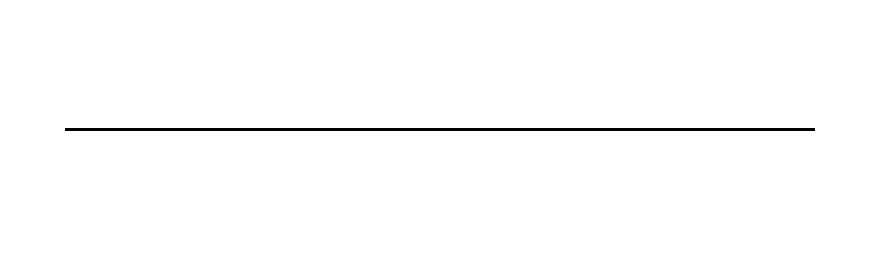}
&
\scalebox{0.7}{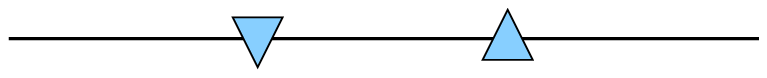}\\
& \\
\end{tabular}
}
\caption{Dual  Vertex Insertion/Deletion Procedure. Left to right: Dual Vertex Insertion; right to left: Dual Vertex Deletion. The edges may be regular or dangling; the oppositely oriented triangluar vertices represent a dual pair of functions.}
\label{fig:coupleInsertDelete}
\end{figure}

\begin{lem}
\label{lem:coupleInsertDelete}
Applying a Dual Vertex Insertion or Deletion Procedure to an NFG preserves the realized exterior function.
\end{lem}
\noindent {\em Proof:}
The Dual
Vertex Insertion Procedure is equivalent to first inserting a $\delta_{=}$ function in the edge and then splitting the $\delta_{=}$ function into a pair of dual functions. The lemma then follows from Lemmas  \ref{lem:vertexGroupSplit} and \ref{lem:eqInsertDelete}. \hfill $\Box$

\subsection{A Linear Algebraic Perspective}
\label{subsec:algebra}

Before we proceed to introduce holographic transformations, we pause to interpret NFGs
from a linear algebraic perspective.

We will denote the set of all complex-valued functions on a finite alphabet ${\cal X}$ by ${\mathbb C}^{\cal X}$. It is well known that ${\mathbb C}^{\cal X}$ is isomorphic to the vector space ${\mathbb C}^{|{\cal X}|}$: after imposing an order on ${\cal X}$, one can arrange the values of any function $f\in {\mathbb C}^{\cal X}$ as a vector  in ${\mathbb C}^{|{\cal X}|}$ according to that order. Similarly, depending  on the structure of ${\cal X}$, the function $f$ may also be viewed as a matrix, or as its higher-dimensional generalization, namely a tensor; if ${\cal X}$ is the cartesian product ${\cal X}_1\times {\cal X}_2$ of some alphabets ${\cal X}_1$ and ${\cal X}_2$, then $f$ may be regarded as a matrix; if ${\cal X}$ is a multifold cartesian product of alphabets, then $f$ may be viewed as a multi-dimensional array, or as a tensor. On the other hand, conventional linear algebraic objects like vectors, matrices and tensors may be alternatively regarded as multivariate functions. In particular, a tensor with $n$ indices may be identified with a multivariate function involving $n$ variables.
From this perspective, any sum-of-products form corresponding to an NFG may be viewed as a linear algebraic expression.


In the simplest case, consider a sum-of-products form $\big\langle f(x_{I}), g(x_{J})\big\rangle$ involving exactly two functions
$f:{\cal X}_I\rightarrow {\mathbb C}$ and $g:{\cal X}_J\rightarrow {\mathbb C}$, where
$I$ and $J$ are two finite index sets, possibly intersecting, and where for every $i\in I\cup J$, ${\cal X}_i$ is an arbitrary finite alphabet. It is straightforward to verify the following propositions:
\begin{itemize}
\item If $I=J\neq \emptyset$, then $\langle f,g\rangle$ is the (non-Hermitian) vector inner product, or dot product, $f\cdot g$, where
 $f$ and $g$ are regarded as  $|{\cal X}_I|$-dimensional vectors and ``$\cdot$'' is a dot product.
\item If $I\supset J\neq \emptyset$, then $\langle f,g\rangle$ is the matrix-vector product $f\cdot g$, where $f$ is regarded
as a $|{\cal X}_{I\setminus J}|\times|{\cal X}_{J}|$ matrix, $g$ is regarded as a $|{\cal X}_J|$-dimensional vector, and ``$\cdot$'' is a matrix-vector product.
\item If $I\setminus J$, $J\setminus I$ and $I\cap J$ are all non-empty, then $\langle f,g\rangle$ is the matrix-matrix product $f\cdot g$, where $f$ is regarded as a $|{\cal X}_{I\setminus J}|\times|{\cal X}_{I\cap J}|$ matrix, $g$ is regarded as $|{\cal X}_{I\cap J}|\times |{\cal X}_{J\setminus I}|$ matrix, and ``$\cdot$'' is a matrix-matrix product.

\item If $I$ and $J$ are disjoint and both non-empty, then $\langle f,g\rangle$
 is the vector outer product, matrix Kronecker product, or tensor product, $f\cdot g$, where
 $f$ and $g$ are regarded as two vectors, two matrices, or two tensors, respectively, and ``$\cdot$'' is the corresponding product operation.

 \end{itemize}
 
In summary, this simple sum-of-products form, namely $\langle f,g\rangle$, unifies various notions of ``product" in linear algebra. This unification illustrates the convenience of understanding linear algebraic objects such as vectors, matrices and tensors as multivariate functions, since in this perspective one never needs to be concerned with
whether a vector is a row or column vector, whether a matrix is transposed, and so forth.

A general sum-of-products form which 
involves multiple functions and which can be represented by an NFG may be viewed as a linear algebraic expression involving various such linear algebraic objects and various such notions of product. The fact that the sum-of-products form $\langle \cdot , \cdot , \ldots , \cdot\rangle$ does not depend on how its arguments are ordered contrasts with the standard order-dependent notations in linear algebra.

In this perspective, Lemma \ref{lem:vertexGroupSplit} follows from the order-independent nature of sum-of-products forms, and Lemma \ref{lem:eqInsertDelete} follows from the fact that $\delta_{=}$ is essentially an identity matrix.

In linear algebra, vectors, matrices and tensors may be viewed alternatively as linear maps, which are characterized by the spaces they act on and the product operation used in defining the maps. Similar perspectives can be made explicit in the NFG context. For example, a complex-valued bivariate function $f(x,y)$ defined on ${\cal X}\times {\cal Y}$ may be viewed as two maps: when participating in the sum-of-products form $\big\langle f(x,y), g(y)\big\rangle$ with a function $g:{\cal Y}\rightarrow {\mathbb C}$, $f$ can be viewed as a linear map from the vector space ${\mathbb C}^{\cal Y}$ to the vector space ${\mathbb C}^{\cal X}$; when participating in the sum-of-products form $\big\langle f(x,y), h(x)\big\rangle$ with  a function $h:{\cal X}\rightarrow {\mathbb C}$, $f$ can be viewed as a linear map from the vector space ${\mathbb C}^{\cal X}$ to the vector space ${\mathbb C}^{\cal Y}$.

This aspect allows us to interpret dual functions in two different ways. Suppose that $\Phi:{\cal X}\times {\cal Y} \rightarrow {\mathbb C}$ and $\widehat{\Phi}:{\cal X}\times {\cal Y} \rightarrow {\mathbb C}$ are a pair of dual functions. On one hand, we may view $\Phi$ as a map from ${\mathbb C}^{\cal X}$ to  ${\mathbb C}^{\cal Y}$ and $\widehat{\Phi}$ as a map from ${\mathbb C}^{\cal Y}$ back to  ${\mathbb C}^{\cal X}$. In this view, the composition map $\widehat{\Phi}\circ\Phi$ is the identity map from ${\mathbb C}^{\cal X}$ to ${\mathbb C}^{\cal X}$, and $\widehat{\Phi}$ is essentially the inverse or pseudo-inverse of $\Phi$. On the other hand, we may view $\Phi$ as a map from ${\mathbb C}^{\cal X}$ to  ${\mathbb C}^{\cal Y}$ and $\widehat{\Phi}$ also as a map from ${\mathbb C}^{\cal X}$ to  ${\mathbb C}^{\cal Y}$. In this view, the dot product of two vectors $f$ and $g$ in ${\mathbb C}^{\cal X}$ is preserved after they are mapped, respectively, to vectors $\langle f, \Phi\rangle$ and
$\langle g, \widehat{\Phi}\rangle$ in ${\mathbb C}^{\cal Y}$. This view is justified by
$\langle f,g\rangle
= \left\langle f,  \Phi, \widehat{\Phi}, g\right\rangle
= \left\langle \langle f, \Phi\rangle , \langle \widehat{\Phi}, g\rangle \right\rangle$.

Finally, we justify the term ``transformer'' that was introduced in Section \ref{subsec:procedures}. Suppose that 
the functions $\Phi$ and $\widehat{\Phi}$ on ${\cal X}\times {\cal Y}$ are dual with respect to alphabet ${\cal Y}$, and that
${\cal X}$ and ${\cal Y}$ have the same cardinality. Then we may identify $\Phi$ with its square-matrix representation in which 
the rows are indexed by ${\cal X}$ and the columns are indexed by ${\cal Y}$; similarly, we may identify $\widehat{\Phi}$
with its square-matrix representation in which the rows are indexed by ${\cal Y}$ and the columns are indexed by ${\cal X}$.  The fact that $\Phi$ and $\widehat{\Phi}$ are dual with respect to ${\cal Y}$ implies that the matrix product $\Phi\cdot \widehat{\Phi}$ is the identity matrix. Thus the matrix $\widehat{\Phi}$ is the unique inverse of the matrix $\Phi$, and {\em vice versa}. Therefore, the functions $\Phi$ and $\widehat{\Phi}$ may be regarded as a pair of {\em transformations} (or transformation kernels) that are inverse to each other.

\subsection{Holographic Transformations and Generalized Holant Theorem}
\label{subsec:GHT}

Now we are ready to define holographic transformations.

Suppose that $I$ is a finite index set and that for each $i\in I$, there are two finite alphabets ${\cal X}_i$ and ${\cal Y}_i$ having the same cardinality. We will call a function $\Phi:{\cal X}_I\times {\cal Y}_I\rightarrow {\mathbb C}$ a {\em separable transformation} if $\Phi$ is a transformation from ${\mathbb C}^{{\cal X}_I}$ to ${\mathbb C}^{{\cal Y}_I}$
(namely, there exists a unique function $\widehat{\Phi}:{\cal X}_I\times {\cal Y}_I\rightarrow {\mathbb C}$ such
that $\langle \Phi(x, y), \widehat{\Phi}(x', y)\rangle = \delta_{=}(x, x')$ for all $x, x'\in {\cal X}$), and there exists a
collection of functions $\{\Phi_i\in {\mathbb C}^{{\cal X}_i\times {\cal Y}_i}:i\in I\}$ such that
$\Phi=\prod\limits_{i\in I}\Phi_i$.
Noting that $\prod\limits_{i\in I}\Phi_i$ may be identified with the sum-of-products form $\langle \Phi_i:i\in I\rangle$, we see that transforming any function in $f\in {\mathbb C}^{{\cal X}_I}$ by $\Phi$ is equivalent to evaluating the sum-of-products form
$\big\langle f, \langle \Phi_i:i\in I\rangle\big\rangle$, which can be performed via {\em separately} transforming $f$ by each of the $\Phi_i$'s in an arbitrary order; hence the term ``separable."

It is easy to verify that if $\Phi:=\prod\limits_{i\in I}\Phi_i$ is a separable transformation, then its inverse transformation $\widehat{\Phi}:{\cal X}_I\times {\cal Y}_I\rightarrow {\mathbb C}$ is $\prod\limits_{i\in I}\widehat{\Phi}_i$
where each $\widehat{\Phi}_i: {\cal X}_i\times {\cal Y}_i\rightarrow {\mathbb C}$ is the inverse transformation of $\Phi_i$. It follows that if a transformation is  separable, then so is its inverse.

\vspace{0.3cm}
\noindent {\bf Holographic Transformation} Let ${\cal G}=(V, E^{\rm int}, E^{\rm ext}, f_V)$ be an NFG where for each edge $e\in E:=
E^{\rm ext}\cup E^{\rm int}$, ${\cal X}_e$ is the alphabet of the represented variable. For each $e\in E$, let ${\cal Y}_e$ be another alphabet having the same cardinality as ${\cal X}_e$. For every vertex $v$ and every edge $e\in E(v)$, we associate a transformation
$\Phi_{v,e}:{\cal X}_e\times {\cal Y}_e\rightarrow {\mathbb C}$ such that 
if $e$ is a regular edge connecting vertices $u$ and $v$, then $\Phi_{u, e}$ and $\Phi_{v, e}$ are the inverse transformations of each other.
Let $\Phi_{v}:=\prod\limits_{e\in E(v)} \Phi_{v, e}$  for every vertex $v$. Locally transform each function $f_v$ to the function $F_v\in {\mathbb C}^{{\cal Y}_{E(v)}}$ via $F_v:= \langle f_v, \Phi_v\rangle$, and collectively denote $\{F_v:v\in V\}$ by $F_V$.
We call the NFG ${\cal G}^H:=(V, E^{\rm int}, E^{\rm ext}, F_V)$ the holographic transformation of ${\cal G}$ with respect to the collection of local separable transformations $\{\Phi_v:v\in V\}$. 

\begin{figure}
\centerline{
\begin{tabular}{S@{\hspace{2cm}}S}
\scalebox{0.5}{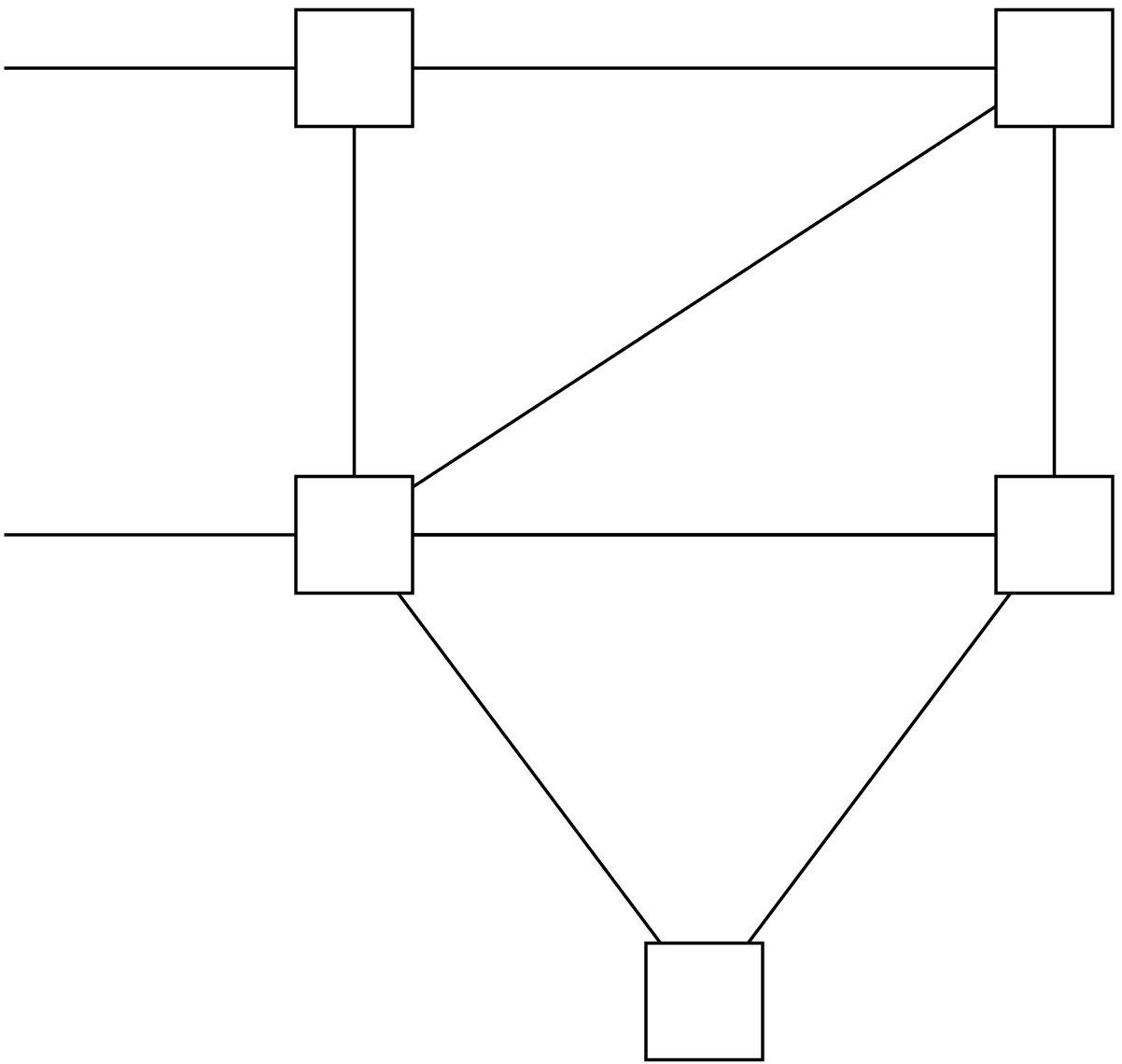}
&
\scalebox{0.5}{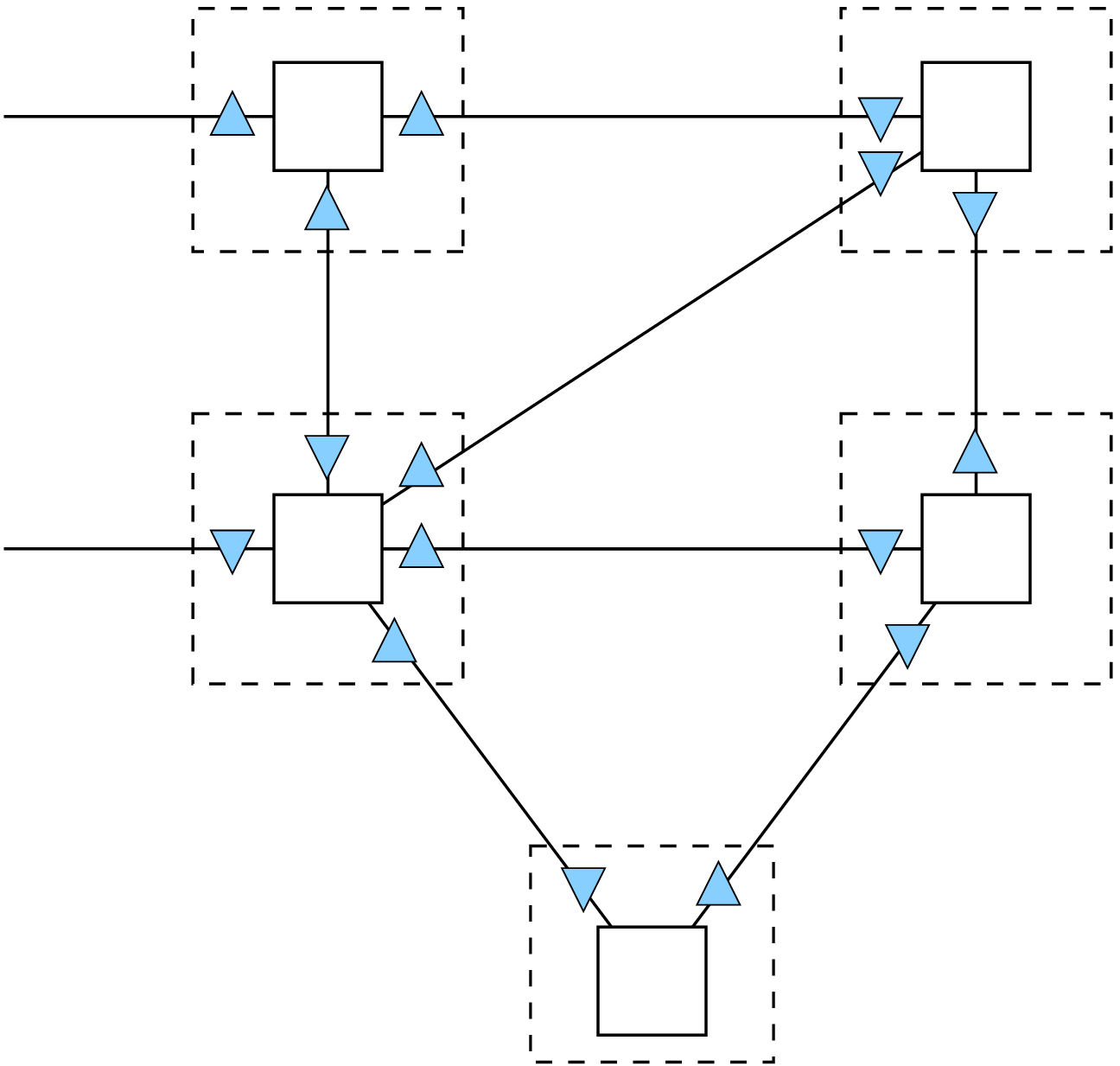}\\
& \\
\end{tabular}
}
\caption{Holographic transformation: an NFG ${\cal G}$ (left) and 
its holographic transformation ${\cal G}^H$ (right). Each triangular vertex is a transformer, possibly different. Oppositely oriented transformers on an edge are the inverses of each other.}
\label{fig:GHT}
\end{figure}

A graphical example of holographic transformation is shown in Figure \ref{fig:GHT}. We note that holographic transformations keep the topology of the NFG unchanged, and only transform each local function.

\begin{thm}[Generalized Holant Theorem] In the setting above, the exterior function $Z_{{\cal G}^H}$ of the NFG ${\cal G}^H$ is related to the exterior function $Z_{{\cal G}}$ of the original NFG ${\cal G}$ by
\[
Z_{{\cal G}^H}(y_{E^{\rm ext}})= \left\langle Z_{{\cal G}}(x_{E^{\rm ext}}), \langle \Phi_e(x_e, y_e): e\in E^{\rm ext}\rangle \right\rangle,
\]
where for each $e\in E^{\rm ext}$, we have written $\Phi_{e}$ in place of $\Phi_{v,e}$.
\end{thm}

\noindent
\emph{Proof:}
The theorem simply follows from Lemma \ref{lem:coupleInsertDelete}. Graphically, as shown in Figure \ref{fig:GHT}, the holographic transformation is equivalent to first inserting into each edge an inverse pair of transformers and into each dangling edge a transformer, and then transforming each local function by its surrounding transformers. Since each inverse pair of transformers cancels out, the only difference between the exterior function of ${\cal G}$ and that of ${\cal G}^H$ is due to the transformers that have been inserted in the dangling edges.  This establishes the theorem. \hfill $\Box$

When $E^{\rm ext}$ is the empty set, the exterior functions of ${\cal G}$ and ${\cal G}^H$ reduce to scalars. In this case, the
generalized Holant theorem reduces to the Holant theorem of \cite{Valiant2004:Holographic}. 

We note also that  in the literature of ``loop calculus'' \cite{Chertkov:Loop1, Chertkov:Loop2, Chertkov:Loop3, Chertkov:easy}, which has been introduced for the study of belief propagation
and of the partition functions of statistical mechanics models, a result equivalent to the Holant theorem has been proved, and a transformation equivalent to our holographic transformation (on NFGs without dangling edges) has been proposed under the name of ``gauge transformation'' (see, \emph{e.g.}, \cite{Chertkov:easy}).

Although our generalization of the Holant theorem appears straightforward, we believe that there is a conceptual leap in this generalization. In particular, the original Holant theorem reveals only that there are redundant and structurally identical NFGs that may be used to represent the same scalar quantity as a sum of products; it makes no attempt to transform or reparameterize an 
exterior function that assumes a more general form. In the general setting of holographic transformations,
when the original NFG ${\cal G}$ is viewed as a realization of some function $g:=Z_{\cal G}$ on a collection of alphabets
$\{{\cal X}_e: e\in E\}$, the holographically transformed NFG ${\cal G}^H$ is viewed as a realization of a related function ${g}^H:=Z_{{\cal G}^H}$ on a different collection of alphabets $\{{\cal Y}_e: e\in E\}$. In particular,
 the function ${g}^H$ may be regarded as a transform-domain representation of $g$ via an ``external change of basis;'' namely, a change of basis for the vector space ${\mathbb C}^{|{\cal X}_{E^{\rm ext}}|}$, that is characterized by the ``external transformation'' $\langle \Phi_e: e\in E^{\rm ext}\rangle$,
where this ``external change of basis'' involves, in its sum-of-products form, a ``local change of basis'' for each component vector space ${\mathbb C}^{|{\cal X}_e|}, e\in E$.

\section{Applications of Holographic Transformations}
\label{sec:application}

\subsection{Duality in Normal Factor Graphs }
\label{sec:forneyDuality}

The first duality theorem for codes on graphs was the normal graph duality theorem that was introduced by Forney in
\cite{Forney2001:Normal}. In the setting of \cite{Forney2001:Normal}, the graphs considered, rather than being NFGs, are ``normal graphs,'' where edges incident on one or two vertices represent ``symbol'' variables and ``state'' variables, respectively, and where each vertex represents a local group-code constraint. The global behavior represented by the graph is the set of all symbol-state configurations that satisfy all local constraints, and  the graph itself represents a group code that consists of all symbol configurations that participate in at least one symbol-state configuration in the global behavior.  In \cite{Forney2001:Normal}, Forney introduced a local ``dualization'' procedure for normal graphs, which converts each local code constraint to its dual code constraint and inserts a ``sign inverter'' into every edge connecting two vertices.
The normal graph duality theorem then states that the dualized graph represents the dual code. 

The normal graph duality theorem of \cite{Forney2001:Normal} may be formulated as an equivalent theorem, which we call the ``code normal factor graph duality theorem,'' using  the language of normal factor graphs. More specifically, 
we may use an NFG to represent a state realization of a code $C$, where each vertex represents the indicator function of a local code constraint, and the exterior function is, up to scale,\footnote{The scaling constant is the number of symbol-state configurations in the global behavior that correspond to each codeword;  this number is the same for every codeword since the global behavior is an abelian group.} the indicator function of the code $C$. The dualization procedure may be reformulated on the NFG as converting the indicator function of each local code to the indicator function of the dual of the local code and inserting an indicator function $\delta_{+}$ into each edge, where $\delta_{+}$ evaluates to $1$ if and only if the two arguments of the function are additive inverses of each other. Then the code normal factor graph duality theorem states that the exterior function realized by the dual NFG is up to scale the indicator function of the dual code $C^\perp$.

In the framework of factor graphs,  Mao and Kschischang \cite{Mao2005:FGFT} introduced the notions of multivariate convolution and convolutional factor graphs, and proved a duality theorem between a multiplicative factor graph and its dual convolutional factor graph. The duality theorem of  \cite{Mao2005:FGFT} (Theorem 11), which we call the ``MK theorem,'' states that a dual pair of  factor graphs represent a Fourier transform pair. Since the indicator function of a code and that of its dual code are a Fourier transform pair up to scale, the code normal graph duality theorem and hence the normal graph duality theorem follow from the  MK theorem as corollaries.

In a concurrent development \cite{Forney:MacWilliams2}, Forney has established a general normal factor graph duality theorem,  where the vertices of an NFG can represent arbitrary functions and the dualization procedure is defined as converting each local function to its Fourier transform and inserting a $\delta_{+}$ function into each edge. The general normal factor graph duality theorem states that a dual pair of NFG's represent a Fourier transform pair up to scale. This theorem reduces to the code normal factor graph duality theorem (and hence to the normal graph duality theorem) if each graph vertex is the indicator function of a local code.

In this subsection, we will show that the general normal factor graph duality theorem follows directly from the generalized Holant theorem.

Let ${\cal G}=(V, E^{\rm int}, E^{\rm ext}, f_V)$ be an arbitrary NFG, where each variable alphabet ${\cal X}_e, e\in E=E^{\rm ext}\cup E^{\rm  int}$, is a finite abelian group written additively. It is well-known that every finite abelian group $({\cal X}, +)$ has a character group ${\cal X}^\wedge$, consisting of precisely the set of all homomorphisms, called characters, of ${\cal X}$ mapping ${\cal X}$ into the multiplicative group of the unit circle in the complex plane. The character group ${\cal X}^\wedge$ of ${\cal X}$ has the following properties \cite{Forney:transformGroups}.
\begin{itemize}
\item  The group operation $+$ in ${\cal X}^\wedge$ is defined by $(\hat{x}_1+\hat{x}_2)(x)=\hat{x}_1(x)\hat{x}_2(x)$
for any two characters $\hat{x}_1, \hat{x}_2\in {\cal X}^\wedge$ and any $x\in {\cal X}$.
\item $\left({\cal X}^\wedge\right)^\wedge$ is isomorphic to ${\cal X}$. This result, known as Pontryagin duality \cite{Forney:transformGroups},  allows  each element of ${\cal X}$ to be treated as a character of ${\cal X}^\wedge$.

\item ${\cal X}^\wedge$ is isomorphic to ${\cal X}$.
\item For each $x\in {\cal X}$ and $\hat{x}\in {\cal X}^\wedge$, $x(\hat{x})=\hat{x}(x)$.  We will denote\footnote{It is customary in the literature to denote both $x(\hat{x})$ and $\hat{x}(x)$ by the pairing $\langle x, \hat{x}\rangle$. But we choose not to use this notation since it collides with our notation for ``sum-of-products'' forms.} both
$x(\hat{x})$ and $\hat{x}(x)$ by $\kappa_{\cal X}(x, \hat{x})$ and, for later use, denote
$\kappa_{\cal X}(x, -\hat{x})/|{\cal X}|$ by $\widehat{\kappa}_{\cal X}(x, \hat{x})$. Keeping in mind that $\kappa_{\cal X}$
and $\widehat{\kappa}_{\cal X}$
 are both defined with respect to the alphabet ${\cal X}$, we may sometimes suppress such dependency in our notation.
It is easy to see that $\kappa_{\cal X}$ and $\widehat{\kappa}_{\cal X}$ are a dual pair of functions (with respect to either
${\cal X}$ or ${\cal X}^\wedge$). Since ${\cal X}$ and ${\cal X}^\wedge$ have the same size,  they in fact define a pair of transformations, namely, the 
Fourier transform and its inverse, as we state next. 

\item For any function $f\in {\mathbb C}^{\cal X}$, its Fourier transform ${\cal F}[f]$ is a complex-valued function on ${\cal X}^\wedge$ defined by
   ${\cal F}[f]:= \langle f, \kappa \rangle.$
    It follows that for any function $f\in {\mathbb C}^{{\cal X}^\wedge}$, its inverse Fourier transform ${\cal F}^{-1}[f]$ is a complex-valued function  on ${\cal X}$,
    ${\cal F}^{-1}[f]
    = \langle f,\widehat{\kappa}\rangle.
    $
    We note that the inverse Fourier transform operator ${\cal F}^{-1}$ may also be applied to a function $f\in {\mathbb C}^{\cal X}$ and result in
    a function on ${\cal X}^\wedge$, where in the sum-of-products form the summation is over the ${\cal X}$-valued variable.
\item If ${\cal X}$ is the direct product ${\cal X}_1\times {\cal X}_2$ of finite abelian groups ${\cal X}_1$ and ${\cal X}_2$, then ${\cal X}^\wedge$ is the direct product ${\cal X}^\wedge_1\times {\cal X}^\wedge_2$ of the character groups ${\cal X}^\wedge_1$ and ${\cal X}^\wedge_2$. In this case, for any $(x_1, x_2)\in {\cal X}_1\times {\cal X}_2$ and any
    $(\hat{x}_1, \hat{x}_2)\in {\cal X}^\wedge_1\times {\cal X}^\wedge_2$,
    \begin{equation}
    \label{eq:ftSeparable}
    \kappa\big( (x_1, x_2), (\hat{x}_1, \hat{x}_2)\big) = \kappa(x_1, \hat{x}_1)\kappa(x_2, \hat{x}_2).
    \end{equation}
Equation (\ref{eq:ftSeparable}) shows that the Fourier transform is a separable transformation.
\end{itemize}

Returning to the NFG ${\cal G}$, we next obtain its ``dual'' by applying Forney's dualization procedure.

\vspace{0.5cm}

\noindent {\bf NFG Dualization Procedure}
Replace each ${\cal X}_e$-valued variable $x_e$ by an ${\cal X}^\wedge_e$-valued the variable $\hat{x}_e$. Replace each function $f_v, v\in V,$ by its Fourier transform ${\cal F}[f_v]$ and insert into each internal edge a vertex representing the function $\delta_{+}(\cdot)$. Again, the function $\delta_{+}$ is an indicator function which evaluates to $1$ if
and only if its two variables are the additive inverses of each other. We will denote the resulting NFG by $\widehat{\cal G}$, and refer to it as the {\em dual} NFG of ${\cal G}$.

\vspace{0.5cm}

\begin{thm}[General Normal Factor Graph Duality Theorem]
\label{thm:duality}
The exterior function $Z_{\widehat{G}}$ realized by the dual NFG $\widehat{\cal G}$ and the exterior function $Z_{\cal G}$ realized by the original NFG ${\cal G}$ are related by
\begin{equation}
\label{eq:forneyDualFourier}
Z_{\widehat{G}}=|{\cal X}_{E^{\bf \rm int}}|\cdot {\cal F}[Z_{\cal G}].
\end{equation}
\end{thm}

\begin{figure}
\centerline{
\begin{tabular}{S@{\hspace{2cm}}S}
\scalebox{0.5}{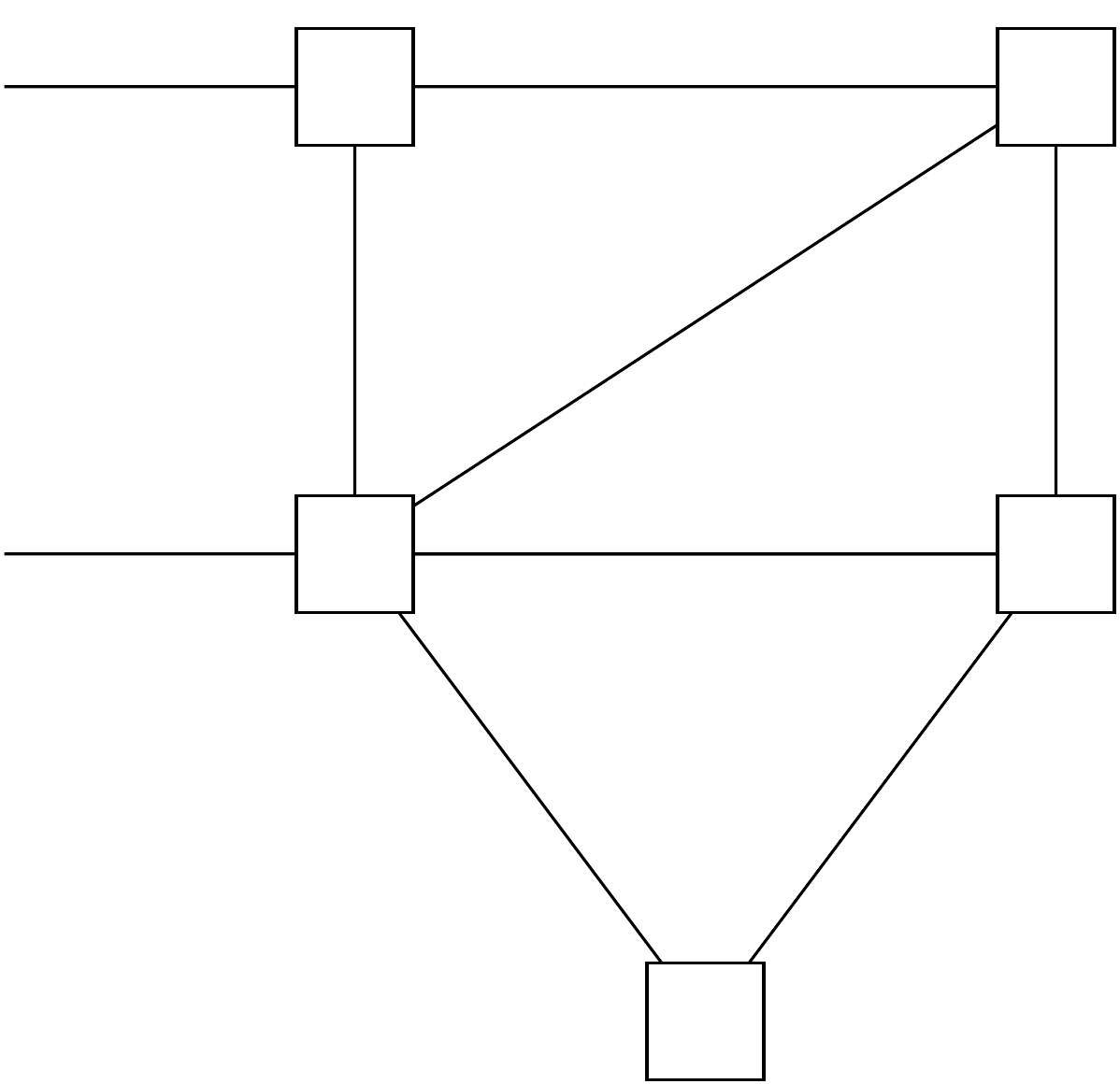}
&
\scalebox{0.5}{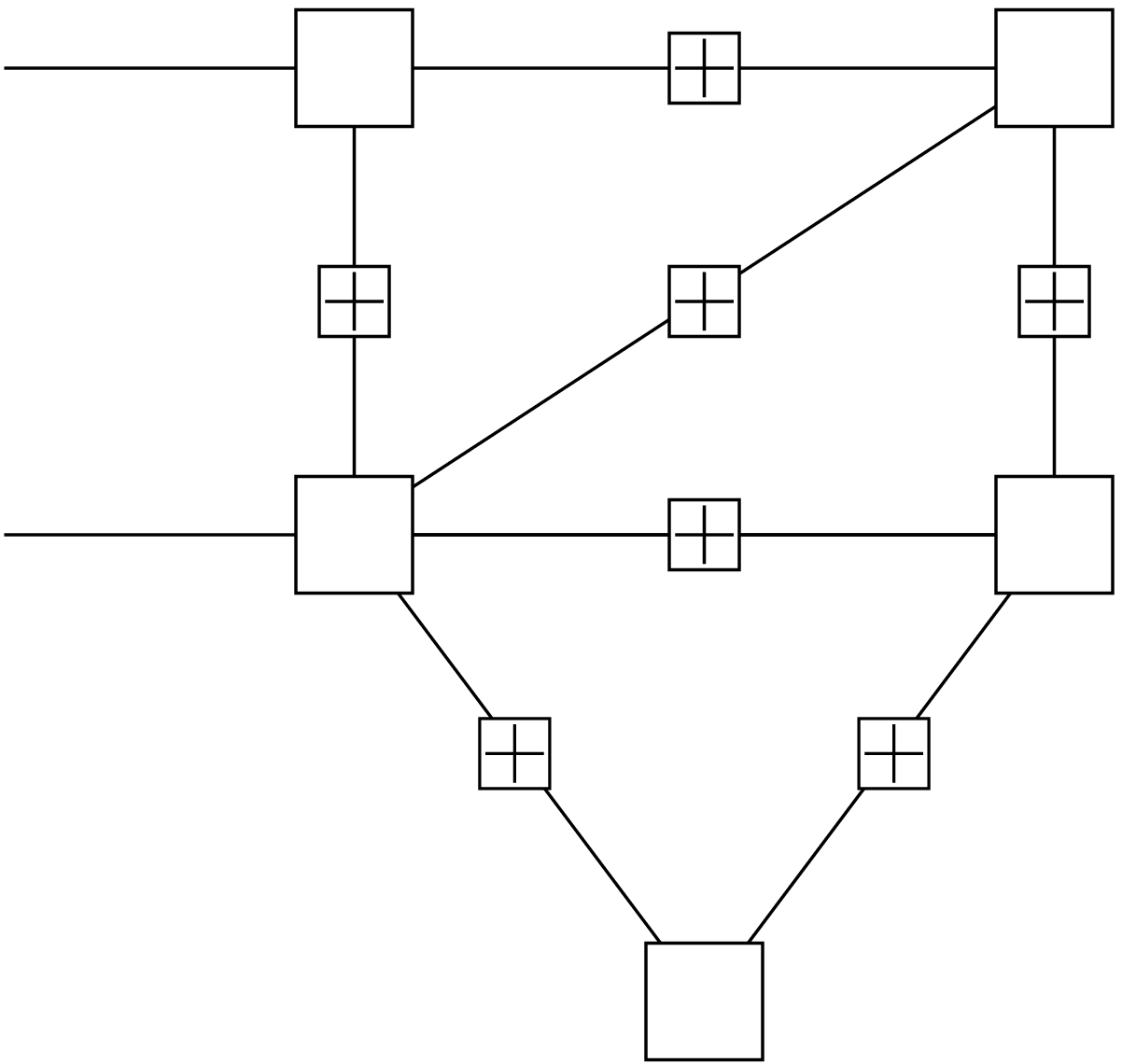}\\
& \\
\scalebox{0.5}{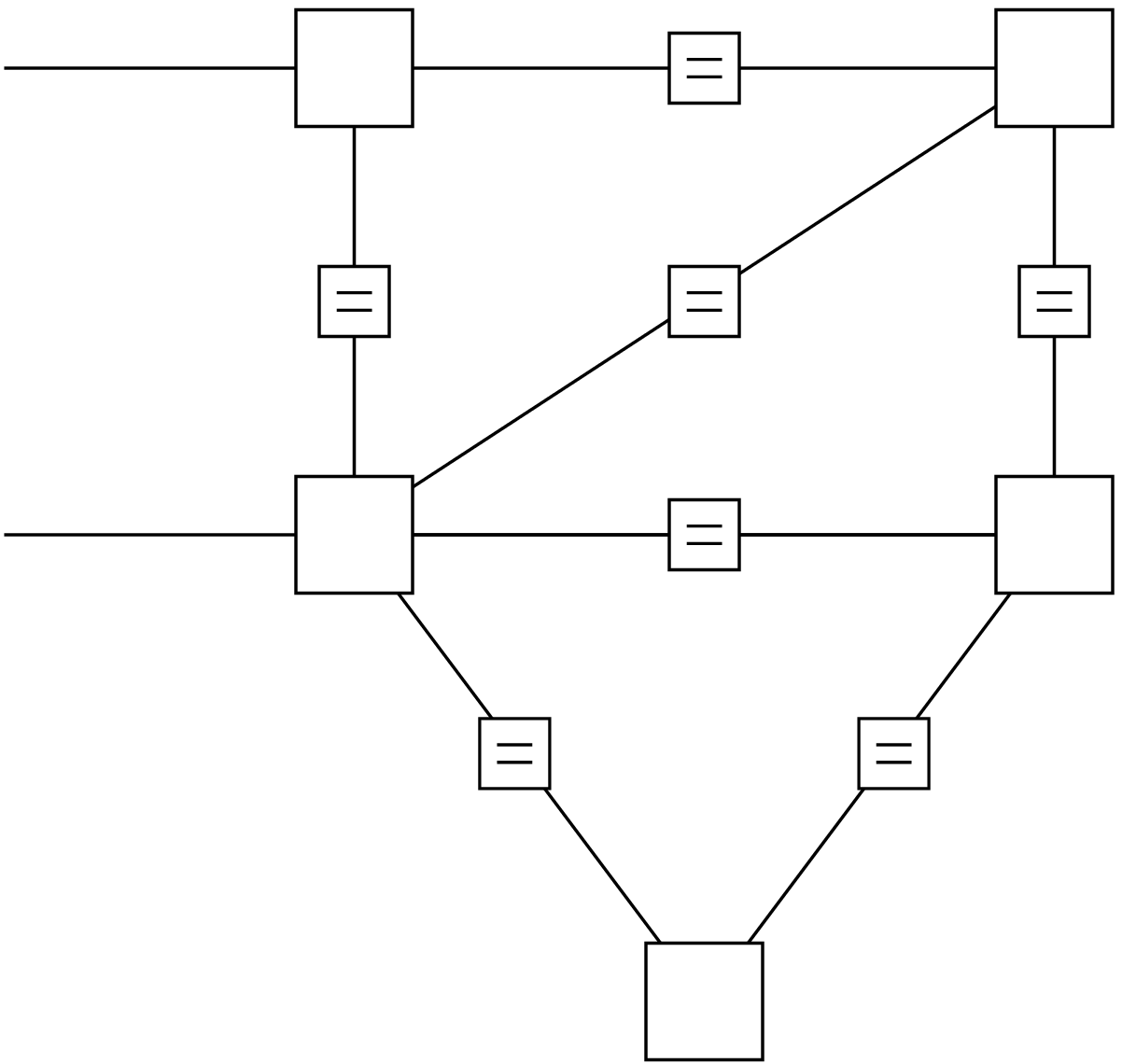}
&
\scalebox{0.5}{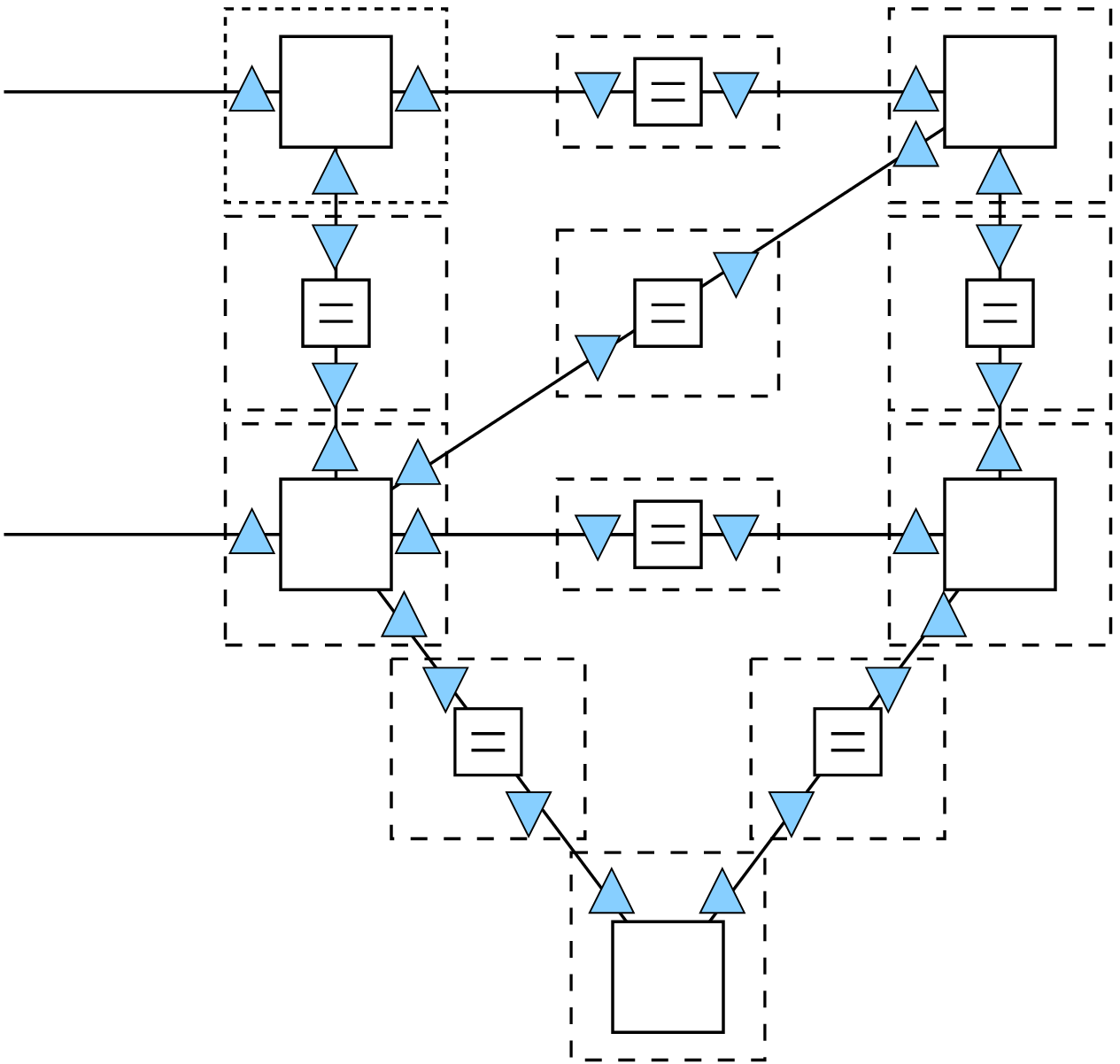}\\
& \\
\end{tabular}
}
\caption{NFG's ${\cal G}$ (top left), $\widehat{\cal G}$ (top right), ${\cal G}'$ (bottom left) and ${{\cal G}'}^H$ (bottom right).}
\label{fig:dualityProof}
\end{figure}

\noindent{\em Proof:} This theorem can be viewed as a corollary of the generalized Holant theorem, and can be simply proved graphically (Figure \ref{fig:dualityProof}):
Construct another NFG ${\cal G}'$ by inserting a  $\delta_{=}$ function into each regular edge ${\cal G}$; by Lemma \ref{lem:eqInsertDelete}, $Z_{{\cal G}'}=Z_{\cal G}$. Obtain the NFG ${{\cal G}'}^H$ from ${\cal G}'$ by
 inverse Fourier transforming every $\delta_{=}$ in ${\cal G}'$
and Fourier transforming every other function.
This corresponds to inserting the dual functions $\kappa_{{\cal X}_e}$ and $\widehat{\kappa}_{{\cal X}_e}$ into each regular edge $e$ (with $\widehat{\kappa}_{{\cal X}_e}$ adjacent to the function $\delta_{=}$) and inserting the function $\kappa_{{\cal X}_e}$ into each dangling edge $e$. Since the inserted transformers
in each regular edge are the inverses of each other, this verifies that ${{\cal G}'}^H$ is a holographic transformation of ${\cal G}'$.
By the generalized Holant theorem,
\[Z_{{{\cal G}'}^H}=\big\langle Z_{{\cal G}'}, \langle \kappa_{e}: e\in E^{\rm  ext}\rangle \big\rangle
=\big\langle Z_{{\cal G}}, \langle \kappa_{e}: e\in E^{\rm  ext}\rangle \big\rangle
={\cal F}[Z_{\cal G}].
\]
Invoking a well-known result that for $\delta_{=}$ defined on ${\cal X}\times {\cal X}$,
${\cal F}^{-1}[\delta_{=}]
=\frac{1}{|{\cal X}|}\delta_{+},
$
we see that ${{\cal G}'}^H$ and $\widehat{\cal G}$ are in fact identical except that in ${{\cal G}'}^H$, each $\delta_{+}$ inserted in edge $e$ is
scaled by $\frac{1}{|{\cal X}_e|}$. The theorem is then proved by collecting all the scaling factors. \hfill $\Box$

We note that Theorem \ref{thm:duality} is the most general NFG duality theorem. If each function in an NFG is an indicator function of a local code, then the exterior function realized by the NFG is up to scale the indicator function of a group code.
Such an NFG, which may be called a ``code normal factor graph'' (``code NFG'') may then be used to represent the group code. This 
 makes a code NFG equivalent to a normal graph. 
Since the indicator function of a code and that of its dual are (up to scale) a Fourier transform pair, the general normal factor graph duality theorem then reduces to the code normal graph duality theorem (and the normal graph duality theorem), which state that a dual pair of code NFGs (resp.\ a dual pair of 
normal graphs) represent a pair of dual codes.

It is worth noting that the general normal factor graph duality theorem also follows from the MK theorem, via the application of the ``projection-slice theorem'' of the Fourier transform, or the ``sampling/averaging duality theorem'' of \cite[Theorem 8]{Mao2005:FGFT}. 
However, the proof using the generalized Holant 
theorem seems more transparent.



\subsection{Holographic Reduction}
\label{subsec:reduction}

A holographic reduction may be regarded as a particular kind of holographic transformation applied to an NFG without dangling edges, by which a counting problem may be reduced to an equivalent ``PerfMatch problem''. Using this technique, Valiant 
constructed polynomial-time solvers for various families of ``counting'' problems previously unknown to be in P \cite{Valiant2004:Holographic}. We now summarize the main results of Valiant \cite{Valiant2004:Holographic} and explain how  holographic reduction works.

\vspace{0.5cm}

\noindent {\bf The PerfMatch Problem}   Suppose that $H = (V,E, w)$ is a weighted graph with vertex set $V$, edge set $E$, and weighting  function $w$ which assigns to each edge $e\in E$ a complex weight $w(e)$. The quantity PerfMatch $\pi(H)$ of $H$ is defined as
\[
\pi(H): = \sum_{M \in Q(H)} \prod_{e \in M} w(e),
\]
where $Q(H)$ is the collection of all perfect matchings\footnote{In graph theory, a perfect matching of a graph is a set of non-adjacent edges such that every vertex of the graph is the endpoint of an edge in the set.} of $H$. It is known that the PerfMatch problem, namely, solving for $\pi(H)$, can be performed in  polynomial time using the FKT algorithm \cite{Kasteleyn, Temperley-Fisher} if $H$ is a \emph{planar} graph.

\vspace{0.5cm}

\begin{figure}
\centerline{
\begin{tabular}{c@{\hspace{1cm}}c}
\scalebox{0.5}{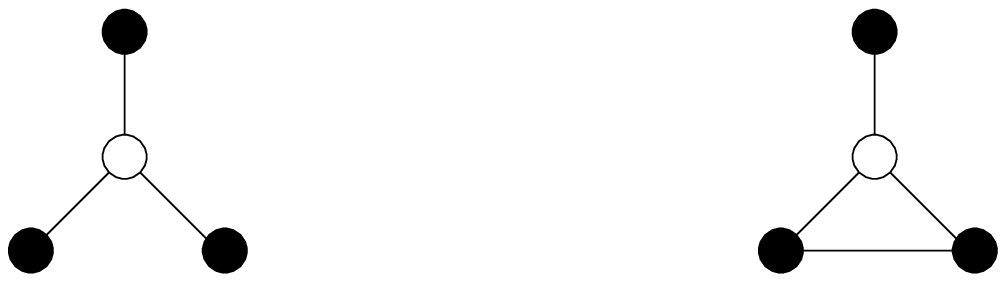} & \scalebox{0.5 }{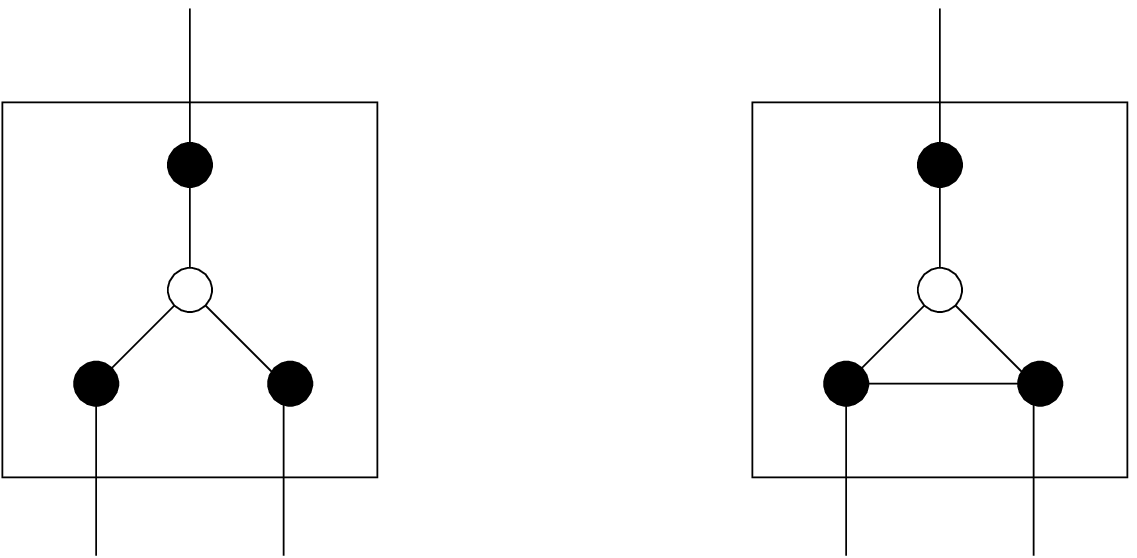}
\\
(a) & (b)\\[20pt]
\scalebox{0.45}{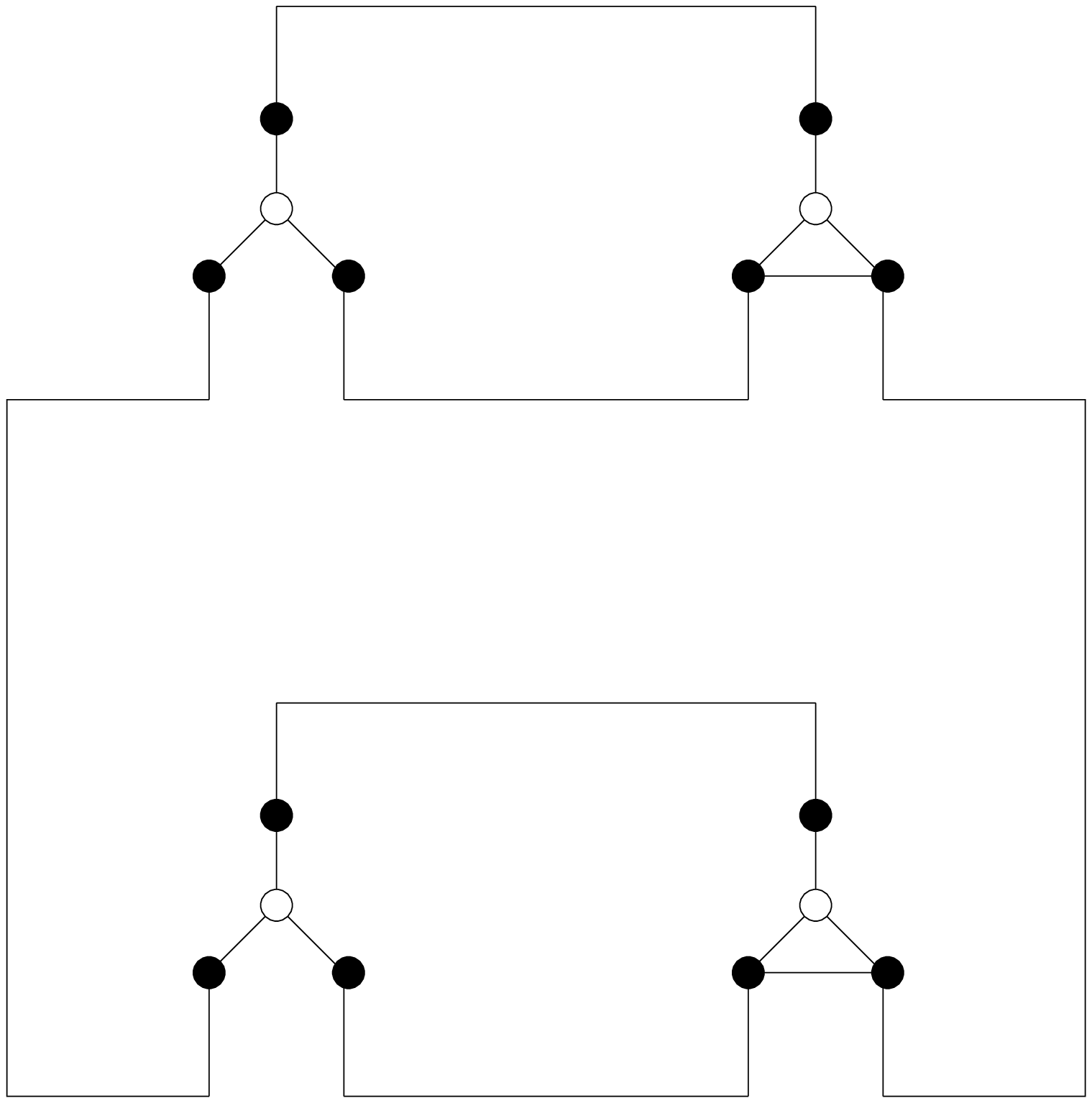}&
\scalebox{0.45}{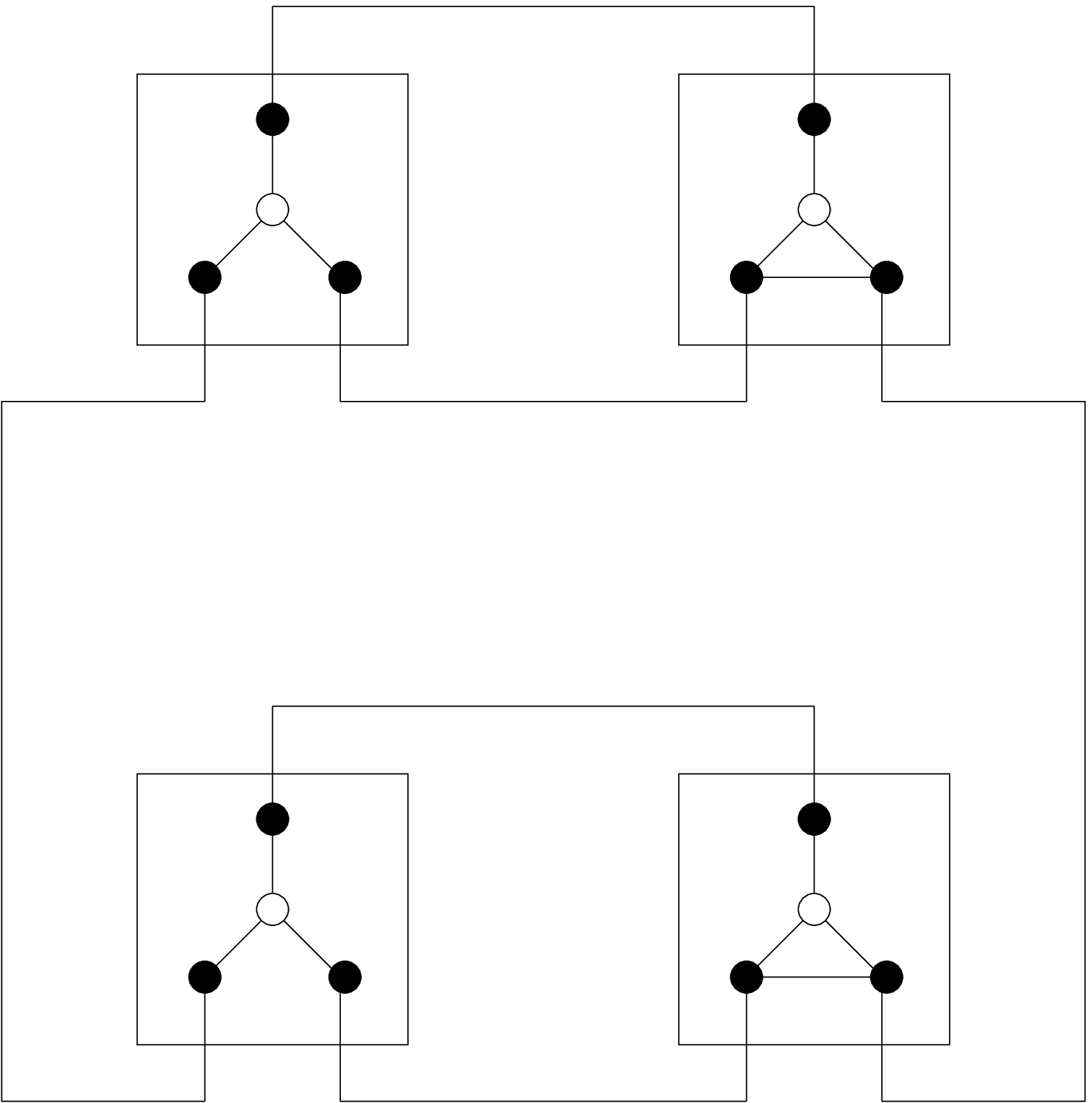}\\
(c) & (d)
\end{tabular}
}
\caption{
\label{fig:gates}
(a) Two matchgates, where solid circles represent the external vertices; (b) the signatures of the matchgates represented as NFG function vertices (the boxes); (c) a larger graph $H$ constructed from the matchgates; (d) the
NFG realization of the PerfMatch of $H$, where each box is a function vertex.}
\end{figure}

A principle underlying holographic reduction is a graph-theoretic property which expresses the PerfMatch of a weighted graph
as a sum-of-products form, in which each involved function is defined based on a local component of the graph. Such a local component is referred to as a ``matchgate,'' and each involved function is referred to as the ``signature'' of such a matchgate.
More precisely, a matchgate is a weighted graph $H$ (which will be used as a local component of a larger graph) with
 a subset $W$ of its vertices specified as its ``external vertices'' (which will be used to connect to other matchgates to form a
 larger graph). Suppose that $(H_1, W_1), (H_2, W_2), \ldots, (H_m, W_m)$  are a collection of matchgates. We may build a larger graph $H$ by connecting these matchgates via their external vertices where the only restriction is that each external vertex of a matchgate $(H_i, W_i)$  connects to exactly one external vertex of a different matchgate $(H_j, W_j)$. The edges that connect the matchgates will be assigned weight $1$. Figure \ref{fig:gates} (a) shows two kinds of matchgates, which are used to construct a larger weighted graph as shown in Figure \ref{fig:gates} (c) in such a way.

 Now let ${\cal E}$ denote the set of edges in the larger graph $H$ that connect (the external vertices of) the matchgates, and let ${\cal E}(i)$ denote the subset of the edges in ${\cal E}$ incident to the external vertices of the matchgate $(H_i, W_i)$.
 Associate with each $e\in {\cal E}$ a $\{0, 1\}$-valued variable $x_e$. The signature $\mu_i$ of the matchgate $(H_i, W_i)$
 is a function of the variable set $x_{{\cal E}(i)}$ defined as follows:  Every configuration $x_{{\cal E}(i)}$ is made to correspond to a subgraph of $H_i$ induced by deleting a subset of its external vertices; more precisely, an external vertex is deleted if and only if it is the endpoint of an edge $e\in {\cal E}(i)$ whose $x_e=1$ in the configuration $x_{{\cal E}(i)}$; the PerfMatch of the subgraph induced this way is then defined to be the value $\mu_i(x_{{\cal E}(i)})$. Then it is possible to express the PerfMatch of the larger graph $H$ as a sum-of-products form involving the  signatures of the matchgates as follows.
\begin{equation}
\label{eq:decomposePerfMatch}
\pi(H)=\sum_{x_{\cal E}} \prod_{i=1}^m \mu_{i}(x_{{\cal E}(i)}).
\end{equation}
It is easy to verify that the sum-of-products form in (\ref{eq:decomposePerfMatch}) has an NFG representation, since each variable
$x_e, e\in {\cal E},$ is involved in precisely two functions (noting that every edge $e\in {\cal E}$ connects two matchgates). In this case, the NFG has no dangling edges, and the realized exterior function reduces to a scalar, i.e., the PerfMatch of the larger graph $H$. 

Figure \ref{fig:gates} shows an example of how to build an NFG that realizes the PerfMatch of a graph using the signatures of its matchgates.
In the figure, (a) shows two matchgates,  where the signature of each matchgate by itself can be viewed as an NFG vertex in (b). When we use the matchgates in  (a) to build the larger graph $H$ in (c), then equality  (\ref{eq:decomposePerfMatch}) suggests that the PerfMatch of  $H$ is realized by the NFG in (d). That is, the NFG topology is identical to the topology by which the matchgates form the larger graph $H$. Visually, the relationship between the NFG and the graph $H$ is apparent: The picture in Figure \ref{fig:gates}(d) is the NFG if we ignore the details inside the boxes, and is the graph $H$ if we ignore the boxes.

\vspace{0.5cm}

\noindent{\bf Solving Counting Problems via Holographic Reduction}
Many counting problems are described in terms of a large collection of variables and a large collection of constraints each involving a subset of the variables. The objective of such problems is to compute the total number of global variable configurations satisfying all the constraints. In this context, the idea of holographic reduction  is to transform the problem of interest to a PerfMatch problem. We now outline this approach.

\begin{enumerate}
\item Express the problem as the computation of the exterior function realized by a {\em planar} NFG ${\cal G}$ without dangling edges. When this is  possible, each 
vertex of ${\cal G}$ represents an indicator (\emph{i.e}., $\{0, 1\}$-valued) function.
\item For each variable $x_e$ in ${\cal G}$, find a pair of inverse transformations $\Phi_e$ and $\widehat{\Phi}_e$, and construct
 a holographic transformation ${\cal G}^H$ of ${\cal G}$ such that each function vertex in ${\cal G}^H$ represents the signature of a matchgate.
\item Create a weighted graph $H$ by replacing each vertex of ${\cal G}^H$ with the corresponding matchgate drawing, and assign weight one to each edge of ${\cal G}^H$.
    This process essentially turns the holographically transformed NFG as in Figure \ref{fig:gates}(d) into a weighted graph as in Figure \ref{fig:gates}(c).
\end{enumerate}

By the Holant theorem, the exterior function  realized by ${\cal G}$ is the same as that realized by ${\cal G}^H$. Since ${\cal G}$ and ${\cal G}^{H}$ do not 
have dangling edges, the realized exterior function is in fact a scalar; by (\ref{eq:decomposePerfMatch}), this scalar is the PerfMatch of $H$. It is easy to verify that ${\cal G}$ being a planar graph implies that $H$ is a planar graph (provided that each matchgate is also a planar graph). Thus solving the PerfMatch problem  for $H$ via the FKT algorithm solves the
 the original counting problem in polynomial time.

 In \cite{Valiant2004:Holographic}, finding the ``right'' transformations $\{\Phi_e: e\in E\}$ that transform each local function in the original NFG to the signature of some matchgate appeared to be an art. Later  Cai and Lu \cite{Cai2007:Art} presented a  systematic approach to determine whether such  a transformation exists.
 Remarkably, Cai {\em et al.\ }\cite{Cai2008:Fibonacci} have extended the approach
of holographic reduction beyond transformations to the PerfMatch problem by introducing the
concept of ``Fibonacci gates."

\section{Concluding Remarks}
\label{sec:conclude}

Sums of products are fundamental in physics, computer science, coding theory, information
theory, and indeed all of science and applied mathematics. In this paper, we have introduced
what we call the ``exterior-function semantics'' for normal factor graphs, which establishes
a one-to-one correspondence between sum-of-products forms satisfying certain nonrestrictive
``normal'' constraints and their associated normal factor graphs. Within this framework, we have introduced 
a very general notion of holographic transformations of normal
factor graphs, and have stated and proved a general and powerful theorem (which we call
the generalized Holant theorem) that relates the exterior function of a normal factor graph
to that of its holographic transformation. As corollaries of this theorem, we obtain Valiant's
original Holant theorem, as well as a very simple proof of a general normal factor graph duality theorem, of which Forney's original normal graph duality theorem is a special case. This
connection between two seemingly distant fields seems to us remarkable.

Although the use of internal variables in graphical models is by no means new, the exterior-function semantics introduced in this paper seems to us elegant, intuitive, and potentially
of wide application. The linear algebraic perspectives that we have mentioned briefly in this
paper may have much more general use. Indeed, as Pascal Vontobel has observed \cite{ForneyPascal:Private}, the
use of ``trace diagrams'' in mathematical physics (see, e.g., \cite{Peterson2009:1, Peterson:Cayley-Hamilton}) appears to have much in common with our graphical techniques. We suspect that the areas of potential applications
are vast.

\section*{Acknowledgment}

 We would like to thank Frank Kschischang for introducing to us holographic algorithms and for earlier discussions on related subjects. We also want to thank the anonymous reviewers for their helpful suggestions. We are particularly indebted to Pascal Vontobel and David Forney for their extensive and very detailed 
comments on previous drafts of this paper, which have helped significantly to improve the  presentation of this paper.

\section*{Appendix: Converting Arbitrary Sum-of-Products Forms to NFGs}

In the framework of factor graphs \cite{frank:factor}, the product of any collection of multivariate functions may be represented by a factor graph. By specifying a subset of the variables in the factor graph to be ``internal'' (namely, to be summed over), it is then possible to represent {\em any} sum-of-products form using a factor graph with additional marks on some variable vertices. 

\begin{figure}
\centerline{
\scalebox{0.5}{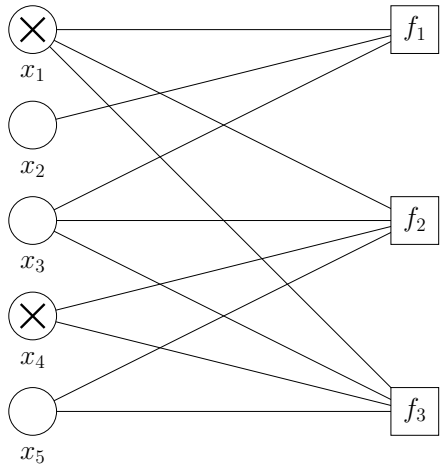}
}
\caption{\label{fig:Z} The ``marked'' factor graph representing 
the sum-of-products form $\sum\limits_{x_1, x_4} f_1(x_1, x_2, x_3)f_2(x_1, x_3, x_4, x_5)f_3(x_1, x_3, x_4, x_5)$.}
\end{figure}

Figure 
\ref{fig:Z} is an example of a sum-of-products form represented by such a ``marked'' factor graph, where the variable vertices marked with ``$\times$'' represent internal variables;  the variable vertices without such marks are external variables, namely, those remaining in the argument of the represented function.  
Such a ``marked'' factor graph then represents the product of all local functions with all internal variables summed over; the function resulting from the summation then clearly involves only the external variables, analogous to the exterior function of an NFG.

Since a factor graph can have an unrestricted topology and an arbitrary subset of its variable verices may be marked, it is possible to represent any sum-of-products form using a ``marked'' factor graph.

Without loss of generality, we will assume that there are no degree-1 internal variable vertices in a ``marked'' factor graph, since otherwise it is always possible to modify the local function connecting to the variable by summing over that variable.

The following procedure, operating on a ``marked'' factor graph representations, ``normalizes'' any sum-of-products form.


\vspace{0.3cm}

\begin{figure}
\centerline{
\begin{tabular}{c}
\scalebox{0.6}{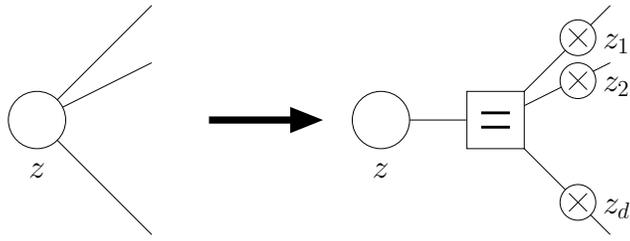}\\
\\
(a)  Replicating an external variable
\\[3pt]
\\
\scalebox{0.6}{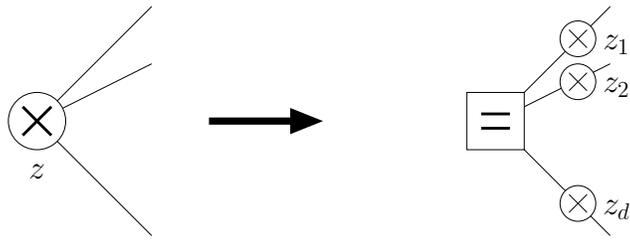}\\
\\
[3pt]
\\
(b)  Replicating an internal variable\\
\\
\end{tabular}
}
\caption{\label{fig:varReplicate} Variable Replication Procedure.}
\end{figure}

\noindent {\bf Variable Replication Procedure}. Suppose that a variable $z$ in a factor graph has degree $d$. Then we may create $d$ replicas $\{z_1, z_2, \ldots, z_d\}$ of variable $z$, isolate $z$ from its edges, and attach each of the $d$ replicas to one of these edges. Remove $z$ if $z$ is an internal variable in the original factor graph. Connect all the replicas of $z$ and $z$ itself, if it is kept, to a new function vertex representing $\delta_{=}(\cdot)$. 
Finally, mark all replicas of $z$ internal (i.e., with ``$\times$''). 
Figure \ref{fig:varReplicate} is a graphical illustration of this procedure.

\vspace{0.5cm}

A procedure similar to the Vertex Replication Procedure above was first 
presented in \cite{Forney2001:Normal}. It is easy to verify that when applying the Variable Replication Procedure to any variable vertex, the sum-of-products form corresponding to the resulting ``marked'' factor graph expresses the same function as the original sum-of-products form does.

On a ``marked'' factor graph, we may apply this procedure to every variable that does not satisfy the ``normal'' degree restriction (namely that an internal variable vertex  have degree two and an external variable vertex have degree one).
It is straightforward to verify that 
in the resulting ``marked'' factor graph, the normal degree
restriction  is necessarily satisfied by all variables. We can then represent the resulting sum-of-products form using the NFG notation, representing internal variables as edges and external variables as dangling edges.

\begin{figure}
\centerline{
\begin{tabular}{c@{\hspace{1cm}}c}
\scalebox{0.45}{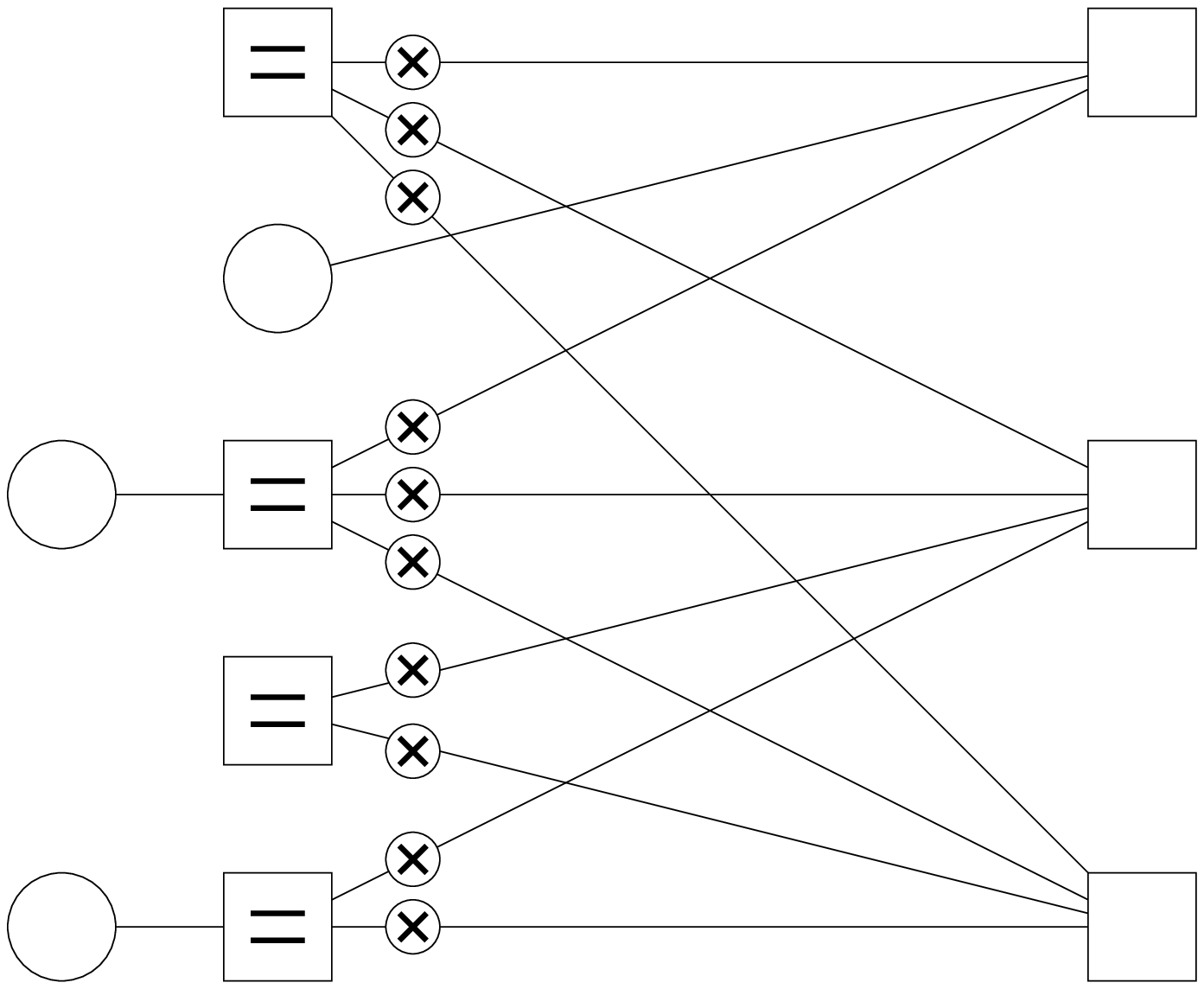} &
\scalebox{0.45}{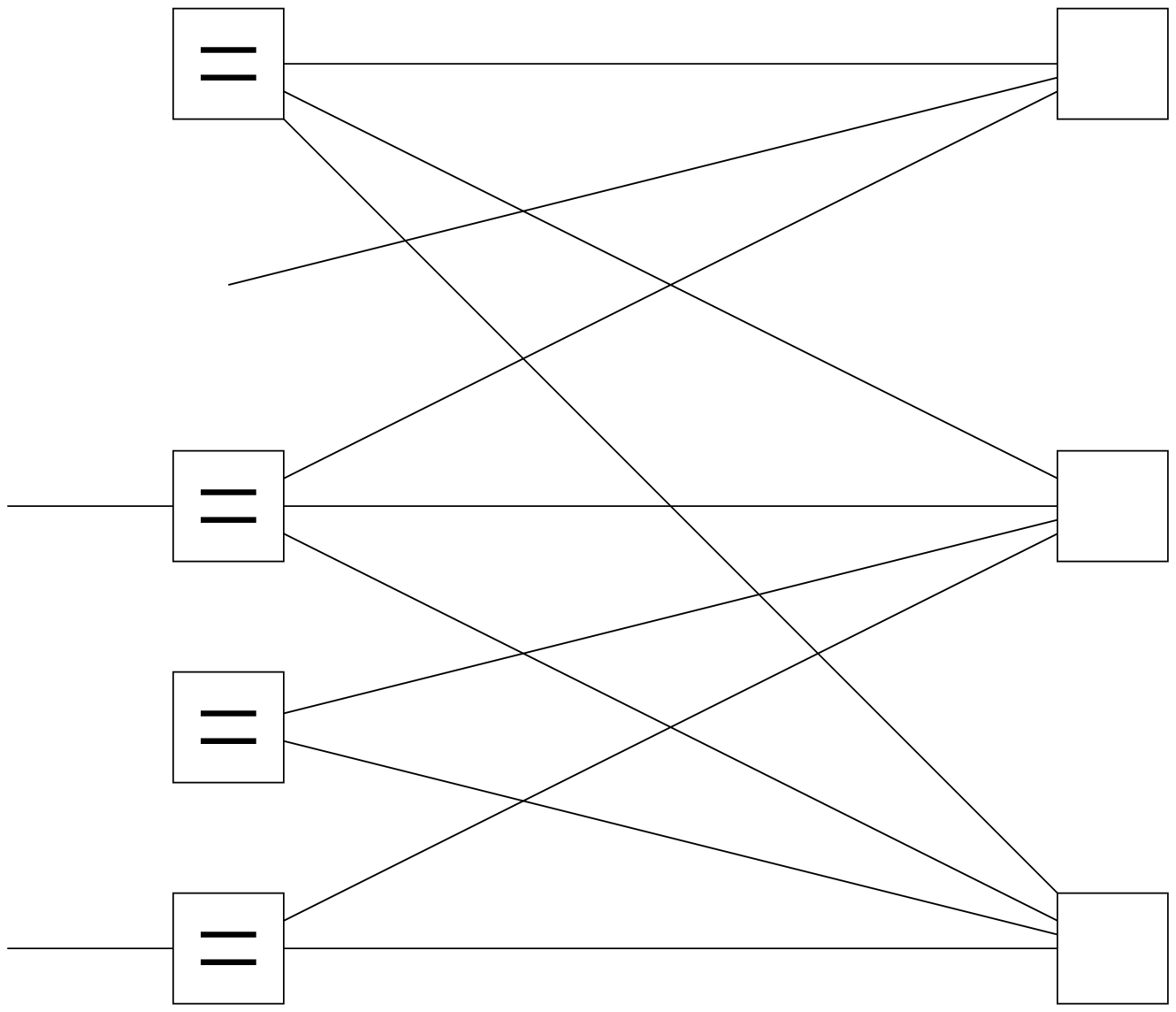}\\
\end{tabular}
}
\caption{\label{fig:FG2NFG} 
The sum-of-products form resulting from normalizing the ``marked'' factor graph
of Figure \ref{fig:Z}. Left: the sum-of-products form represented 
as a ``marked'' factor graph; right: the sum-of-product form represented as an NFG.
}
\end{figure}

Figure \ref{fig:FG2NFG} shows the sum-of-products form resulting from normalizing the ``marked'' factor graph in Figure \ref{fig:Z}.

\bibliography{Holographic_revised}

\end{document}